\documentclass[prd, aps, superscriptaddress, showkeys]{revtex4}
\usepackage{latexsym, graphicx}
\usepackage{epstopdf}
\usepackage{amssymb}

\begin{document}

\title{Entanglement Dynamics between Inertial and Non-uniformly Accelerated Detectors}
\author{David C. M. Ostapchuk}
\email{davidostapchuk@alumni.uwaterloo.ca}
\affiliation{Department of Physics and Astronomy, University of Waterloo,
Waterloo, Canada N2L 3G1}
\author{Shih-Yuin Lin}
\email{sylin@cc.ncue.edu.tw}
\thanks{Corresponding author}
\affiliation{Department of Physics, National Changhua University of Education,
Changhua 50007, Taiwan}
\author{Robert B. Mann}
\email{rbmann@sciborg.uwaterloo.ca}
\affiliation{Department of Physics and Astronomy, University of Waterloo,
Waterloo, Canada N2L 3G1}
\author{B. L. Hu}
\email{blhu@umd.edu}
\affiliation{Maryland Center for Fundamental Physics and Joint Quantum
Institute, University of Maryland, College Park, Maryland 20742-4111, USA}



\date{June 13, 2012}

\begin{abstract}
We study the time-dependence of quantum entanglement 
between two Unruh-DeWitt detectors, one at rest in a Minkowski frame, 
the other non-uniformly accelerated in some specified way. 
The two detectors each couple to a scalar quantum field but do not interact directly. 
The primary challenge in problems involving non-uniformly accelerated detectors arises from the
fact that an event horizon is absent and the Unruh temperature is ill-defined. By
numerical calculation we demonstrate that the correlators  
of the accelerated detector in the weak coupling limit 
behaves like those of an oscillator in a bath of time-varying
``temperature'' proportional to the instantaneous proper acceleration of the
detector, with oscillatory modifications due to non-adiabatic effects. We
find that in this setup the acceleration of the detector in
effect slows down the disentanglement process in Minkowski time due to the time dilation
in that moving detector. 
\end{abstract}


\maketitle

\section{Introduction}

A uniformly accelerated, point-like observer moving in a quantum field in a
Minkowski vacuum will experience the same effect as an inertial observer in a
thermal field at a temperature proportional to the proper acceleration of the
observer. This is called the Unruh effect~\cite{Un76}, and the temperature
experienced by the observer is called the Unruh temperature. Since the observer
is uniformly accelerated and assumed to be point-like, for such an observer one
can sharply define the event horizon, beyond which no information can reach the
observer. The corresponding geometry is the Rindler space where all uniformly
accelerated detectors follow stationary trajectories in its right wedge
R. This model, first proposed by Unruh to understand the Hawking effect in a
black hole, has garnered wide-spread attention on its own merit in a variety of
contexts, relativistic quantum information being one of the most
recent. 

In recent years much effort has been made to understand the quantum
informational aspects of the Unruh effect in various setups and for different
quantum fields~\cite{AM03, SU05, FueMan05, FueMan06, OPhD08, OM09, LCH08, LH09,
LH10}. Most of the work employs arguments that rely on the existence of an
event horizon. For example, one common way to show the thermality experienced
by a uniformly accelerated detector on the right wedge R of the Rindler spacetime is to argue that the
event horizon acts to divide the spacetime into two regions leaving one region,
the left wedge L, totally inaccessible to observers in the right wedge R. Upon
tracing out the field modes in L one sees easily that the observer in R
experiences thermality at the Unruh temperature proportional to its proper
acceleration. Nevertheless, in cases without a Rindler-like spacetime structure
or the presence of an event horizon neither thermality arguments nor geometric
properties are of much use. More generally, in cases where the observer
undergoes non-uniform acceleration there is no timelike  Killing vector nor
equilibrium condition to define the Unruh temperature for all times.

The question of how a detector experiences the effects of a quantum field 
when it undergoes non-uniform acceleration was raised in the 90s by one of the 
present authors~\cite{RHK}: in the absence of an event horizon, will it detect
radiation or not? For a purist endorsing only geometric arguments who insists
on the event horizon being the determining factor, the answer would be no. But
physically there is no fundamental distinction between nonuniform and uniform
acceleration, and one would expect radiation, albeit not in a strictly thermal
form. In fact it is natural to ask how the detector responds to a change in
kinematic states, say, from an inertial state to a uniformly accelerated state,
or the reverse.  This is a more generic case --- what one encounters in the
everyday experience of driving a car to go somewhere and back. This intuitive
view, though simple, is not easy to formulate, because of the non-availability
of an event horizon (from the geometric viewpoint) or an equilibrium condition
(from the field theory viewpoint). For a general trajectory one should treat
the detector-quantum field system under fully nonequilibrium conditions. Using
stochastic field theory (based on the influence functional representation of
the quantum field, which departs from an equilibrium condition as the detector
deviates from  uniform acceleration), the physical predictions mentioned above
(that the observer experience nonthermal radiance) were
confirmed~\cite{JHCapri}. Furthermore, quantum field theory in the influence
functional or in-in (closed-time-path, or Schwinger-Keldysh) formulation, which
is designed to treat causal evolutions (as distinct from the traditional in-out
formulation for scattering problems, or imposing future dynamical conditions),
liberates the physical essence of the problem from the limitations (or utility,
but only when applicable) of geometric constructs such as an event
horizon~\footnote{The event horizon being defined at the infinite future, an
assumption of its existence is rather unnatural for a dynamical
(time-evolutionary) problem as the boundary condition is set in the future, not
in the past, i.e., it is introduced in a teleological way, see,
e.g.,~\cite{Wilczek} for pathologies arising from such setups.}, certainly away
from the gravitation and general relativity context, and attribute the Unruh
effect purely as a kinematic effect related to the excitations by vacuum
fluctuations by the motion of the detector in a quantum field~\footnote{We feel
that the approach to problems of this nature based on quantum field theory and
statistical mechanics, or nonequilibrium quantum field theory (NEqQFT), is
generally more malleable and functional than the geometric approach, which,
like the etiquette of the aristocrats, is invariably more elegant but
restricted. We hasten to add that in contrast to the NEqQFT approach, there
have also been developments along the geometric-spacetime approach to relax the
notion of event horizon to more general situations, from a strictly global to a
more quasi-local sense, such as the isolated horizon of Hayward, Ashtekar et
al~\cite{Hayward, Ash} but the effects of quantum fields in spacetimes with
such constructs have yet to be explored to make comparison with the predictions
of the nonequilibrium field theory results of the 90s.}.
 The kinematic viewpoint and the nonequilibrium approach are clearly more 
encompassing and widely applicable. For example the ``circular Unruh effect" 
\cite{BL83, JHCapri}, which has been related to the Sokorov-Ternov effect observed in storage rings, 
can be regarded as a nonequilibrium QED effect manifest theoretically when 
electrons are treated as point-like objects rather than plane waves in space.  

One way to study the dynamics of a non-uniformly accelerated detector is to look at
the response function in the transition probability of a uniformly accelerated detector with finite coupling
time to the vacuum \cite{SS92, HMP93, Pad96, OP03}. One may argue that in the interaction region 
the detector acts like a uniformly accelerated one, while it behaves like an inertial detector in the asymptotic past,
provided there is no excitation in an inertial detector initially in its ground state. Other
attempts for the off-uniform acceleration cases, e.g. \cite{Sc04, OM07, CV99, KP10}, 
are also focused on the response functions.
We want to issue a note of caution here: the transition probabilities associated with these detector response 
functions are usually considered using time-dependent perturbation theory, which is valid only 
in the weak coupling (transient) limit with a nonvanishing proper acceleration \cite{LH07}. 
Overall, for the study of entanglement dynamics of two detectors 
the response functions are not as convenient as the correlators, which are what we set forth to calculate below. 
From the correlators of two detectors traversing the full history in different states of motion, 
we can extract the entanglement dynamics between them.

Previous work using nonequilibrium quantum field theory (NEqQFT --- for an
introduction, see, e.g.,~\cite{CalHu08}) concentrated primarily on the
fluctuation-dissipation aspects of particles and their energy spectrum.
Here we are interested in the quantum informational aspects of two non-uniformly
accelerated detectors. We are specifically concerned with how quantum
entanglement between these two detectors evolves in time, especially in
comparison to the previously studied cases of (a) two inertial
detectors~\cite{LH09} and (b) between one inertial detector and one
uniformly accelerated detector~\cite{AM03, SU05, FueMan05, LCH08}. 
As a first step we therefore consider the situation in which
one detector $A$ remains at rest and a second detector $B$ starts out from an
inertial state and ends up in a uniformly accelerated state.  We expect this
scenario to be a hybrid nature of cases (a) and (b) above.

There is a coordinate system in Minkowski spacetime that gives a simpler
description of the motion of detector $B$. It is given by
\begin{equation}
  ds^2 = (e^{-2w\xi}+e^{2w\zeta})(-d\xi^2+ d\zeta^2) + dy^2 + dz^2,
\label{CVframe}
\end{equation}
which was first introduced by Kalnins~\cite{Kalnins} and later used by Costa
and Villalba and others~\cite{CV,PV92,VM92} for the analysis of quantum field theory of a
detector undergoing non-uniform acceleration, and more recently used to study
the entanglement degradation due to non-inertial motion
by one of the present authors ~\cite{MV09}. 
The range of the coordinates is $-\infty < (\xi, \zeta) < \infty$, which covers 
half of the Minkowski space (the region $x>t$.) An observer with worldline 
$\zeta = \zeta_0$ for some constant $\zeta_0$ is inertial in the asymptotic
past and has uniform acceleration in the asymptotic future. In this sense the
coordinates~(\ref{CVframe}) resemble Minkowski coordinates as
$t, \xi\to -\infty$ and Rindler coordinates at $t, \xi\to +\infty$.  Note that
since the trajectory asymptotes to one of uniform acceleration we do have
available a single event horizon, in contrast to the two horizons present for a
Rindler observer.

Taking advantage of this simple description, we consider two Unruh-DeWitt (UD)
detectors $A$ and $B$, with $A$ at rest in conventional Minkowski coordinates,
and $B$ accelerated non-uniformly so as to be stationary in the
coordinates~(\ref{CVframe}), in a quantum field initially in the Minkowski
vacuum. The analysis here follows the approach of prior work by two of us on
two UD detectors~\cite{LCH08, LH09}. Since the motions of the two detectors are
highly asymmetric, we have to resort to numerical computation to obtain the results. Conceptually, our
results demonstrate both the methods and the nature of entanglement degradation
in situations without global geometric constructs such as an event horizon.
From this perspective our approach provides a useful case study for comparison.
In addition, our results are useful for a description of the quantum twin paradox 
problem~\cite{LinBehHu}, with the setup depicted in~\cite{RHK} where 
one tries to predict the logbook of entanglement dynamics between these two famous twins, 
one staying home whilst the other travels away and returns. The intellectual question is 
how their entanglement alters upon return in comparison to both twins staying at home, 
and how entanglement in the outbound trip differs from the return trip;  the technical 
difficulty in this situation is that the returning twin does not see an event horizon. 

The paper is organized as follows. In Sec.~\ref{theModel} we introduce the setup of our
model. We show some selected results on the evolution of the self and cross correlators of the  
detectors in Secs.~\ref{SingDet} and \ref{XcorrMt}, and mutual influences of the
detectors are discussed in Sec.~\ref{secmuinf}. Then the entanglement dynamics between
the detectors in Minkowski time will be demonstrated in Sec.~\ref{EntDyn}, and a summary
follows in Sec.~\ref{summa}. In appendix~\ref{retard} we give the retarded distance and the
retarded time between the two detectors, and 
in appendix~\ref{CalcCorr} we include details of the numerical calculations for the
self and  cross correlators  of the detectors. 
Finally in appendix~\ref{jump} we explain some 
interesting behavior  of the self correlators of the non-uniformly accelerated
detector during and after the transition observed in our numerical results. 

\section{A non-uniformly accelerated detector}
\label{theModel}

\begin{figure}
\includegraphics[width=5cm]{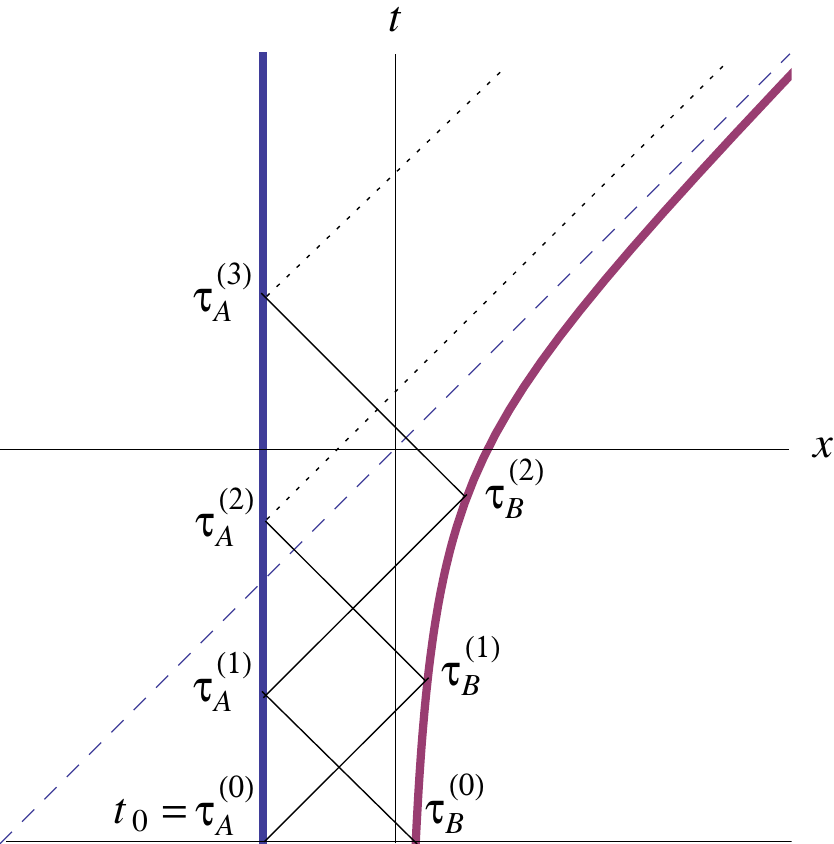}
\caption{The trajectories of the detectors $A$ (left thick curve) and $B$
(right thick curve) in the Minkowski frame. The dashed line in the plot is the
event horizon for detector $B$. $\tau_{A}^{(n)}$ and $\tau_{B}^{(n)}$ denote
the moments that the $n$-th order mutual influences on the detectors $A$ and
$B$ come into play (see section~\protect{\ref{secmuinf}}).}
\label{defmutinf}
\end{figure}

Consider the dynamics of two UD detectors coupled with a quantum field. The
action is given by~\cite{LCH08}
\begin{eqnarray}
  S &=& -\int d^4 x \sqrt{-g} {1\over 2}\partial_\mu\Phi(x) \partial^\mu\Phi(x) + 
  \sum_{{\bf d}=A,B}\int d\tau_{\bf d}\times \nonumber\\ & &
   \left\{ {m_0\over 2}\left[\left(\partial_{\bf d}Q_{\bf d}\right)^2
    -\Omega_{0}^2 Q_{\bf d}^2\right]
    + \lambda_0\int d^4 x 
    Q_{\bf d}(\tau_{\bf d})\Phi (x)\delta^4\left(x^{\mu}-z_{\bf d}^{\mu}(\tau_{\bf d})\right)\right\}
  \label{Stot1}
\end{eqnarray}
where $g_{\mu\nu} = {\rm diag}(-1,1,1,1)$,
$\partial_{\bf d}\equiv \partial/\partial \tau_{\bf d}$, $Q_A$ and $Q_B$ are
the internal degrees of freedom of the point-like detectors $A$ and $B$,
assumed to be two identical harmonic oscillators with the same mass $m_0$, bare
natural frequency $\Omega_0$, and the same local time-resolution. The proper
times for $Q_A$ and $Q_B$ are $\tau_A$ and $\tau_B$, respectively. The scalar
field $\Phi$ is assumed to be massless and $\lambda_0$ is the coupling
constant. Detector $A$ is at rest in a Minkowski frame along the world line
\begin{equation}
  z_A^\mu = (t,-d,0,0)
\label{zA}
\end{equation}
whereas detector $B$ moves with non-uniform acceleration, its world line given
in Minkowski space by the Kalnins coordinate
\begin{equation}
  z_B^\mu = \left( {1\over a}\sinh a\xi -{1\over 2a}e^{-a\xi}, {1\over a}\cosh a\xi
    -{1\over 2a}e^{-a\xi},0,0\right),
\label{zB}
\end{equation}
which is at rest at $\zeta=0$ in the non-inertial frame~(\ref{CVframe}) (see
figure~\ref{defmutinf}). From~(\ref{CVframe}) with $\zeta=0$, $d\zeta=dy=dz=0$
and $w=a$, the proper time of detector $B$ is related to the timelike parameter
$\xi$ by
\begin{equation}
  ds^2=-d\tau^2 = (e^{-2a\xi}+1)( - d\xi^2),
\end{equation}
so
\begin{equation}
  {d\tau\over d\xi} = \sqrt{e^{-2a\xi}+1},
\end{equation}
which implies
\begin{eqnarray}
   \tau(\xi) &=& {1\over a}\sinh^{-1}e^{a\xi}-{1\over a}\sqrt{e^{-2a\xi}+1}\\
   &=& \xi + {1\over a}\ln\left( 1+\sqrt{e^{-2a\xi}+1}\right)-{1\over a}\sqrt{e^{-2a\xi}+1}.
\label{TauOfXi}
\end{eqnarray}
The inverse function $\xi(\tau)$ has no closed form and has to be obtained
numerically by finding the root of $\xi$ in the above equation for a given
$\tau$. The 4-velocity and the 4-acceleration of detector $B$ are,
respectively,
\begin{equation}
  v_B^{\mu} = {dz_B^\mu\over d\tau} = {{dz_B^\mu /d\xi}\over{d\tau /d\xi}} =
  {1\over\sqrt{e^{-2a\xi}+1}}\left( \cosh a\xi + {e^{-a\xi}\over 2},
  \sinh a\xi + {e^{-a\xi}\over 2}, 0,0\right),
\end{equation}
and
\begin{equation}
  a_B^\mu = {dv_B^\mu\over d\tau} = \left({a e^{a \xi}\over 2(e^{-2a\xi}+1)^2},
  {a(e^{a \xi}+2e^{-a\xi})\over 2(e^{-2a\xi}+1)^2},0,0\right).
\end{equation}
So the proper acceleration of the accelerated detector $B$ reads
\begin{equation}
  \alpha^{}_B \equiv \sqrt{a^{}_{B\mu} a_B^\mu} = {a\over (e^{-2a\xi}+1)^{3/2}},
\label{avary}
\end{equation}
which approaches zero at the initial moment $t_0\ll -1$ ($\xi\to -\infty$),
increases to $a/2$ at $t\approx -0.114/a$ ($e^{-2a\xi}=2^{2/3}-1$), and then to
$a$ as $t\gg 1$ ($\xi\to \infty$).

We pause to note that the non-adiabatic behaviour of the response functions of a single
non-uniformly accelerated detector has been previously considered by expanding in powers of 
$\dot{a}/a^2$  \cite{KP10}.   Unfortunately such analysis is not practical here
because our detector $B$ has $\dot{\alpha}^{}_B/\alpha_B^2 = 3 e^{-2a\xi(\tau)}$,
which is much greater than $1$ through the early stage of evolution (when $\xi(\tau)$ is
negatively large and $\alpha_B$ is almost zero), but not that large as the proper
acceleration undergoes a transition from approximately $0$ to $a$ around $t\approx 0$.  It would be interesting to look
at the dynamics of a detector having a transition from one non-zero proper acceleration to another in order
to appropriately   compare our results with these earlier ones in \cite{KP10}.
However, this would divert the focus of the present paper, which is concerned with
the dynamics of entanglement between an inertial and a non-uniformly accelerated detector. 
Suffice it to note that the authors of \cite{KP10} observed that the behaviour of the response functions
with $a\gg \Omega$ is qualitatively different from those with $a\ll \Omega$.
We observe similar  behaviour of the self correlators of detector $B$.

\section{Entanglement dynamics and detector correlators}
\label{DynCorDet}

We consider a situation in which the initial state at $t=t_0$ in the Minkowski
frame is a product state of the Minkowski vacuum of the field $\left| 0_M
\right>$ (which is Gaussian) and the Gaussian two-mode squeezed
state~\cite{LH09}
\begin{eqnarray}
  & &\rho^{}_{AB}(Q_A,P_A,Q_B,P_B) =  {1\over \pi^2\hbar^2}\times\nonumber\\ & &
 \exp -{1\over 2}\left[
  {\beta^2\over\hbar^2}\left( Q_A + Q_B\right)^2 +
  {1\over \alpha^2}\left( Q_A - Q_B\right)^2 +
  {\alpha^2\over\hbar^2}\left( P_A - P_B\right)^2 +
  {1\over \beta^2}\left( P_A + P_B\right)^2 \right]
\label{initGauss}
\end{eqnarray}
of the detectors in Wigner representation. At $t=t_0$, the detectors start to
couple with the quantum field. By virtue of the linearity of the combined
system~(\ref{Stot1}),  the quantum state of the combined system,
and therefore the reduced state of the detectors,
will always be Gaussian and fully determined by
the covariance matrix
\begin{equation}
  {\bf V} = \left( \begin{array}{cc} {\bf v}^{}_{AA} & {\bf v}^{}_{AB} \\
                   {\bf v}^{}_{BA} & {\bf v}^{}_{BB} \end{array}\right)
\label{coVm}
\end{equation}
in which the elements of the $2\times 2$ matrices ${\bf v}_{ij}$, $i,j=A,B$ are
those symmetrized two-point correlators
${\bf v}_{ij}{}^{mn} =\left<\right.{\cal R}_i^m , {\cal R}_j^n \left.\right>
\equiv \left<\right. ({\cal R}_i^m{\cal R}_j^n +{\cal R}_j^n{\cal R}_i^m) \left.\right>/2$
with ${\cal R}_i^m = (Q_i(t), P_i(t))$, $m,n=1,2$.
We thus  have   full information of the reduced state of the detector pair at
each moment once we know the history of all the two-point correlators, from which
the dynamics of entanglement between the detectors can be extracted.
	
Also, by virtue of linearity, the operators of the detectors in
the Heisenberg picture will evolve to a linear combination of all the detector
operators $\hat{Q}_{\bf d}$, $\hat{P}_{\bf d}$ (${\bf d}=A,B$) and the field
operators $\hat{\Phi}_{\bf k}$, $\hat{\Pi}_{\bf k}$ defined at the initial
moment $t_0$ as
\begin{equation}
  \hat{Q}_{\bf d}(\tau_{\bf d}) = \hat{Q}^{\rm D}_{\bf d}(\tau_{\bf d}) +
\hat{Q}^{\rm F}_{\bf d}(\tau_{\bf d}),
\end{equation}
where
\begin{eqnarray}
  \hat{Q}^{\rm D}_{\bf d}(\tau_{\bf d}) &\equiv& \sum_{{\bf d'}=A,B}
    \left[\phi_{\bf d}^{\bf d'}(\tau_{\bf d})\hat{Q}^{}_{\bf d'}
    +\pi_{\bf d}^{\bf d'}(\tau_{\bf d})\hat{P}^{}_{\bf d'}\right],\\
  \hat{Q}^{\rm F}_{\bf d}(\tau_{\bf d}) &\equiv& \int {d^3 k\over (2\pi)^3} \left[
	\phi_{\bf d}^{\bf k}(\tau_{\bf d})\hat{\Phi}^{}_{\bf k}+
	\pi_{\bf d}^{\bf k}(\tau_{\bf d})\hat{\Pi}^{}_{\bf k}\right].
\end{eqnarray}
Here $\phi_{\bf d}(\tau_{\bf d})$, $\pi_{\bf d}(\tau_{\bf d})$ are mode
functions, and we have $\hat{P}_{\bf d}(\tau_{\bf d}) = m_0 \partial_{\bf
d}\hat{Q}_{\bf d}(\tau_{\bf d})$ from~(\ref{Stot1}). Then each symmetrized
two-point correlator of the detectors will split into a sum of the a-part and
the v-part~\cite{LH06} as
\begin{equation}
  \left<\right. {\cal R}_{\bf d}(\tau_{\bf d}) ,{\cal R'}_{\bf d'}(\tau_{\bf d'})  \left.\right> =
  \left<\right. {\cal R}_{\bf d}(\tau_{\bf d}) , {\cal R'}_{\bf d'}(\tau_{\bf d'}) \left.\right>_{\rm a} +
  \left<\right. {\cal R}_{\bf d}(\tau_{\bf d}) , {\cal R'}_{\bf d'}(\tau_{\bf d'})  \left.\right>_{\rm v},
\end{equation}
with ${\cal R},{\cal R'}=P,Q$ and
\begin{eqnarray}
 \left<\right. {\cal R}_{\bf d}(\tau_{\bf d}) , {\cal R'}_{\bf d'}(\tau_{\bf d'}) \left.\right>_{\rm a} &\equiv&
   {1\over 2}{\rm Tr}\left[\,\left({\cal R}^{\rm D}_{\bf d}(\tau_{\bf d}) {\cal R'}^{\rm D}_{\bf d'}(\tau_{\bf d'})
   +{\cal R'}^{\rm D}_{\bf d'}(\tau_{\bf d'}){\cal R}^{\rm D}_{\bf d}(\tau_{\bf d})  \right) \rho^{}_{AB}\right] ,\\
 \left<\right. {\cal R}_{\bf d}(\tau_{\bf d}) , {\cal R'}_{\bf d'}(\tau_{\bf d'}) \left.\right>_{\rm v} &\equiv&
   {1\over 2}\left< 0_M\right|({\cal R}^{\rm F}_{\bf d}(\tau_{\bf d}) {\cal R'}^{\rm F}_{\bf d'}(\tau_{\bf d'}) +
   {\cal R'}^{\rm F}_{\bf d'}(\tau_{\bf d'}){\cal R}^{\rm F}_{\bf d}(\tau_{\bf d}) )\left|0_M\right>.
\end{eqnarray}
The a-part corresponds to the initial state of the detector $(\ref{initGauss})$, while the v-part
corresponds to the response to the field vacuum $\left| 0^{}_M \right>$.
The a-parts of the correlators are relatively easy to obtain in the perturbative regime with
large distance between the detectors. Some examples will be given in Sec.~\ref{EntDyn}.
The calculation of the v-parts, however, is more complicated.
Unlike detectors in uniform acceleration, there is no simple symmetry here to
help in obtaining analytic results. All of our computations will be performed
numerically, even in the weak-coupling regime with mutual influences neglected.

\subsection{Dynamics of single detectors}
\label{SingDet}

The reduced state of a single detector is obtained by tracing out the other detector
in the reduced state of the detector pair. Since the latter is Gaussian, the former
must also be a Gaussian state, which is fully determined by the self correlators of
that detector. 

Neglecting mutual influences between the two detectors, 
the self correlators of the inertial detector $A$ have previously been obtained
in closed form~\cite{LH07}. For the accelerated detector $B$, unfortunately,
there is no analytic expression for its self correlators. The v-part of the
latter can be expressed in 2D integrals as, for example,
\begin{eqnarray}
  & &\left<\right.Q_B(\tau),Q_B(\tau')\left.\right>_{\rm v}
   = {\lambda_0^2\over m_0^2 \Omega^2} \times \nonumber\\ & & \hspace{.5cm}{\rm Re}
  \int_{\tau_0}^\tau d\tilde{\tau} \int_{\tau_0}^{\tau'} d\tilde{\tau'}
   K(\tau - \tilde{\tau}) K(\tau - \tilde{\tau}')
   D^+(z_B^{\mu}(\tilde{\tau}), z_B^{\mu}(\tilde{\tau}')),
\end{eqnarray}
where $\tau, \tau' \ge \tau_0 \equiv \tau(t_0)$,
$K(x)\equiv e^{-\gamma x}\sin\Omega x$ with the coupling strength
$\gamma \equiv \lambda_0^2/8\pi m_0$ and the renormalized natural frequency
$\Omega$ of the detector (see eq. (3.59) in ref.~\cite{BD82} and eq.(59) in~\cite{LH06},
where the $\tau$ in $q^{(-)}$ should be $\tau'$), and
\begin{eqnarray}
& & D^+(z_B^{\mu}(\tilde{\tau}), z_B^{\mu}(\tilde{\tau}')) = \nonumber\\
& & \hspace{.5cm}{\hbar /(2\pi)^2 \over 
   \left|{\bf z}^{}_B(\tilde{\tau}-(i\epsilon/2))-{\bf z}^{}_B(\tilde{\tau}'+(i\epsilon/2))\right|^2- 
   \left[ z^0_B(\tilde{\tau}-(i\epsilon/2))-z^0_B(\tilde{\tau}'+(i\epsilon/2))\right]^2}
\label{WiFn}
\end{eqnarray}
is the positive frequency Wightman function of the massless scalar field.
(Note that $(\ref{WiFn})$ is not exactly the same as the one in eq.(3.59) in ref. \cite{BD82}.
The latter can yield unphysical results.
See Refs. \cite{Sc04, OM07} and appendix A.1 in \cite{LH10} for more details.) 
The above integrand is singular at $\tilde{\tau}=\tilde{\tau}'$ if $\epsilon=0$.
To treat this singularity properly we calculate the quantity
\begin{eqnarray}
  \delta\left<\right.Q_B^2(\tau)\left.\right>_{\rm v} &\equiv&
  \lim_{\tau'\to \tau}\left[\left<\right.Q_B(\tau),Q_B(\tau')\left.\right>_{\rm v}-
  \left<\right.Q_B(\tau),Q_B(\tau')\left.\right>_{{\rm v}(a\to 0)}\right] \nonumber\\ &=&
  {2\gamma\hbar\over\pi m_0 \Omega^2}\int_{\tau_0}^\tau d\tilde{\tau} \int_{\tau_0}^\tau d\tilde{\tau'}
  K(\tau - \tilde{\tau}) K(\tau - \tilde{\tau}') \tilde{f}(\tilde{\tau}, \tilde{\tau}')
\label{dQB2def}
\end{eqnarray}
instead, where
\begin{equation}
  \tilde{f}(\tilde{\tau}, \tilde{\tau}')\equiv
    -{a^2\over 4\left( 1 + e^{-a [\xi(\tilde{\tau})+\xi(\tilde{\tau}')]}\right)
    \sinh^2 {a\over 2}[\xi(\tilde{\tau})- \xi(\tilde{\tau}')]} +
    {1\over \left(\tilde{\tau}-\tilde{\tau}' \right)^2}
 \label{tildefdef}
\end{equation}
(with $\epsilon$ neglected).
This is the deviation from
$\left<\right.Q_B(\tau)Q_B(\tau')\left.\right>_{{\rm v}(a\to 0)}$
for inertial detectors in the Minkowski vacuum, namely, the one for detector
$A$ (eq.~$(A9)$ in~\cite{LH07}) with $t$ replaced by $\tau$.

Now the integrand of~(\ref{dQB2def}) is regular and well controlled because
the divergences in the coincidence limit $\tau'\to \tau$ of this theory
(corresponding to the large constants $\Lambda_1$ and $\Lambda_0$ defined in
ref.~\cite{LH07}, which are reference frame independent  
since they are defined via the proper times of the detectors)
all belong to $\left<\right.Q_B^2(\tau)\left.\right>_{{\rm v}(a\to 0)}$. Indeed,
it is straightforward to verify that $\tilde{f}$ in (\ref{tildefdef}) 
go smoothly to $\alpha^{2}_B (T)/12+ O(\tilde{\tau}-\tilde{\tau}')$, which is
regular as $\tilde{\tau}-\tilde{\tau}'\to 0$. Here
$T\equiv(\tilde{\tau} +\tilde{\tau}')/2$, and the proper acceleration
$\alpha^{}_B (T) = \sqrt{a^{}_{B\mu}(\xi(T)) a_B^\mu(\xi(T))}$ has been given in~(\ref{avary}).

The above integrand is suppressed rapidly when $T$ becomes more and more
negative, meaning that the value of the correlator
$\left<\right.Q_B^2(\tau)\left.\right>_{\rm v}$ will be very close to the value
of $\left<\right.Q_B^2(\tau)\left.\right>_{{\rm v}(a\to 0)}$ at
$\Omega\tau \ll -1$, when the detector is almost at rest in Minkowski frame.
After $T \approx 0$, the absolute value of the integrand becomes significant
around $\tilde{\tau} = \tilde{\tau}'$, so the difference
$\delta\left<\right.Q_B^2(\tau)\left.\right>_{\rm v}$ becomes obvious after
$\tau$ becomes positive.

\begin{figure}
\includegraphics[width=6cm]{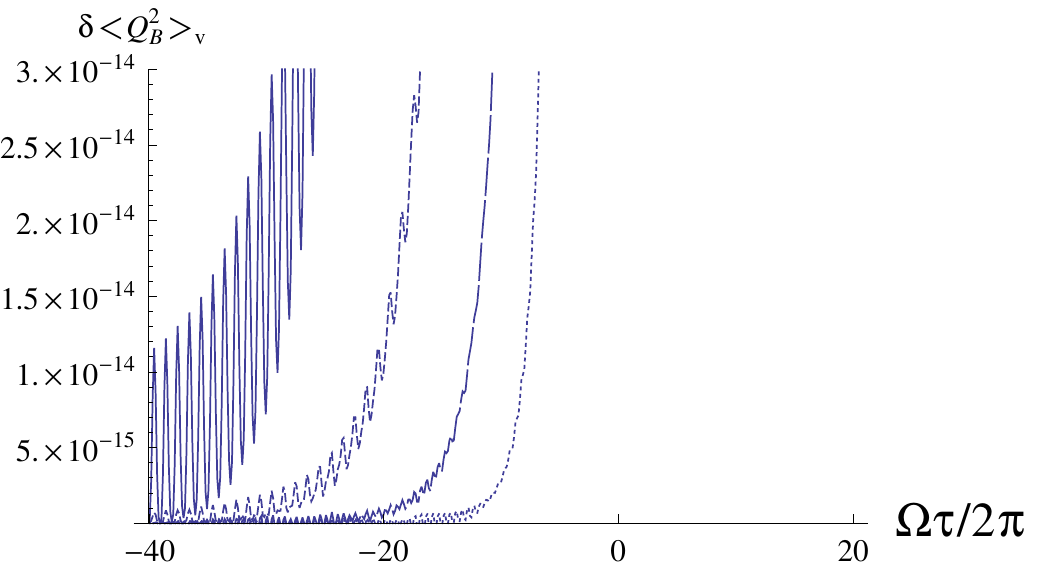}
\includegraphics[width=6cm]{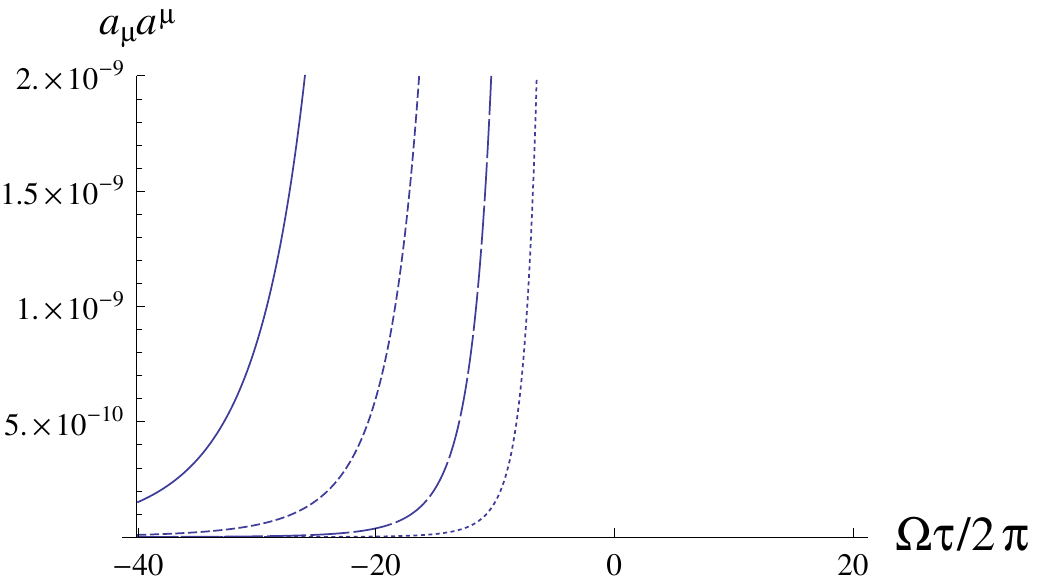}\\
\includegraphics[width=6cm]{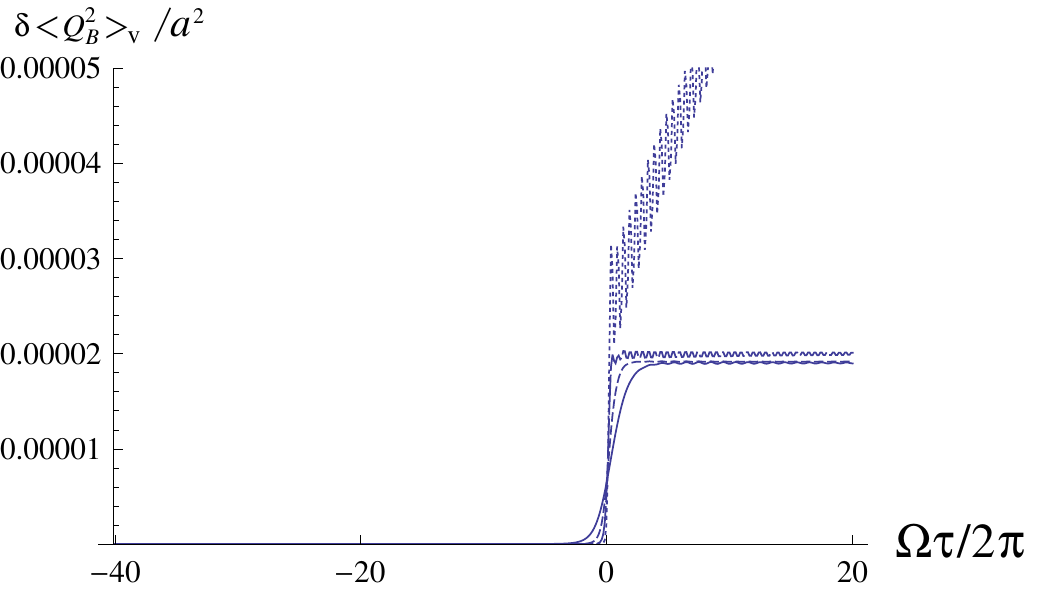}
\includegraphics[width=6cm]{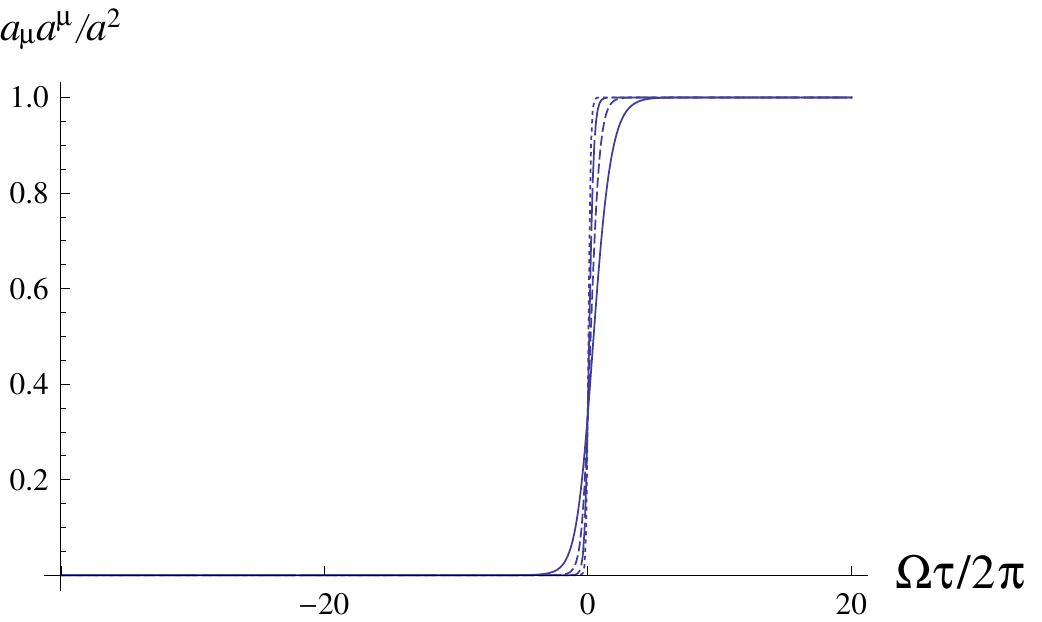}
\caption{Here we take $\gamma=0.01$, $\Omega=2.3$, and $m_0=\hbar=1$. The
curves in these plots are those with $a=1/4$ (solid), $1/2$ (short-dashed),
$1$ (long-dashed), and $2$ (dotted), respectively. (Upper-left) The early
evolution of $\delta\left<\right. Q_B^2(\tau)\left.\right>_{\rm v}$, whose growth
rate is quite similar to the corresponding $\alpha_B^2=a_\mu a^\mu$
(upper-right).
The oscillations at early times are artifacts caused by the impulse
at the initial moment $\tau_0$; they should  vanish as
$\tau_0\to -\infty$ when the proper acceleration is exactly zero. 
(Lower-left) Evolution of $\delta\left<\right. Q_B^2(\tau)\left.\right>_{\rm v}$ normalized by $a^2$.
All curves with $a\le 1$ behave similarly to $\alpha_B^2(\tau)$ (lower-right),
with the small oscillations mainly due to the initial impact at
$\tau = \tau_0= -80\pi/\Omega$.
The oscillations in the curves with $a=2$ are due to a
non-adiabatic effect associated with the small time scale of the rapid growth in acceleration.
Note that here the height of the ``jump'' divided by $a^2$ around $\tau=0$ is
$\gamma/[6\pi(\gamma^2+ \Omega^2)^2] \approx 1.896\times 10^{-5}$
(cf.~(\ref{dQBjump}) with $\hbar=m_0=1$). }
\label{dQB2}
\end{figure}

We show some results in figure~\ref{dQB2} 
(more details on numerical calculations can be found in appendix~\ref{CalcCorr}).
From the lower-left plot of figure~\ref{dQB2} one can see that the value of
$\delta\left<\right.Q^2_B \left.\right>_{\rm v}$ has a ``jump'' around
$\tau \approx 0$ when the proper acceleration significantly departs from zero.
This jump is in fact adiabatic: compared with the lower-right plot of
figure~\ref{dQB2}, the increasing rate of
$\delta\left<\right.Q^2_B \left.\right>_{\rm v}$ during the jump is virtually
the same as the growth rate of $\alpha_B^2$. 
From these numerical results we
observed that the jump around $\tau \approx 0$ is from $0$ to a value about
\begin{equation}
  {\cal Q} \equiv {\gamma \hbar a^2 \over 6\pi m_0(\gamma^2+\Omega^2)^2}.
  \label{dQBjump}
\end{equation}
A discussion on this observation is given in appendix~\ref{jump}.
Note that the difference of the asymptotic values of the two-point functions reads
\begin{eqnarray}
  & &\left.\left<\right.Q^2_B(\infty)\left.\right>_{\rm v}\right|_{a^\mu a_\mu=a^2}- 
  \left.\left<\right.Q^2_B(\infty)\left.\right>_{\rm v}\right|_{a^\mu a_\mu \to 0} \nonumber\\  
  &=&{\hbar \over 2\pi m_0\Omega}\left\{ {\rm Re}\left[{ia\over \gamma+i\Omega}-
    2i\psi\left(1+{\gamma+i\Omega\over a}\right)\right]- i \ln
    {\gamma-i\Omega\over \gamma+i\Omega}\right\} \label{exactasym}\\
&\approx& {\hbar \over 2\pi m_0}{\gamma a^2\over 3(\gamma^2+\Omega^2)^2} +
  O(a^4), \label{apprasym}
\end{eqnarray}
from eqs.~$(A7)$ and~$(A11)$ in~\cite{LH07}, where $\psi(x)$ is the digamma function.
Interestingly enough, the $O(a^2)$ term in~(\ref{apprasym}) is identical to ${\cal Q}$, which is always less
than the value of the left hand side of~(\ref{apprasym}).

After the jump ($\tau \approx 0$), we observe that those values of
$\delta\left<\right.Q_B^2\left.\right>_{\rm v}$ keep growing roughly as
\begin{equation}
  \delta\left<\right.Q^2_B(\tau)\left.\right>_{\rm v} \sim 
  \left[ \left.\left<\right.Q^2_B(\infty)\left.\right>_{\rm v}\right|_{a^\mu a_\mu=a^2}-  
  \left.\left<\right.Q^2_B(\infty)\left.\right>_{\rm v}\right|_{a^\mu a_\mu \to 0}\right](1-e^{-2\gamma\tau}) 
  + {\cal Q} e^{-2\gamma\tau}
  \label{evoweak}
\end{equation}
in the weak coupling limit. This is similar to the behavior of a harmonic
oscillator in contact with a ``thermal'' bath at a time-varying ``temperature''.

\begin{figure}
\includegraphics[width=4.9cm]{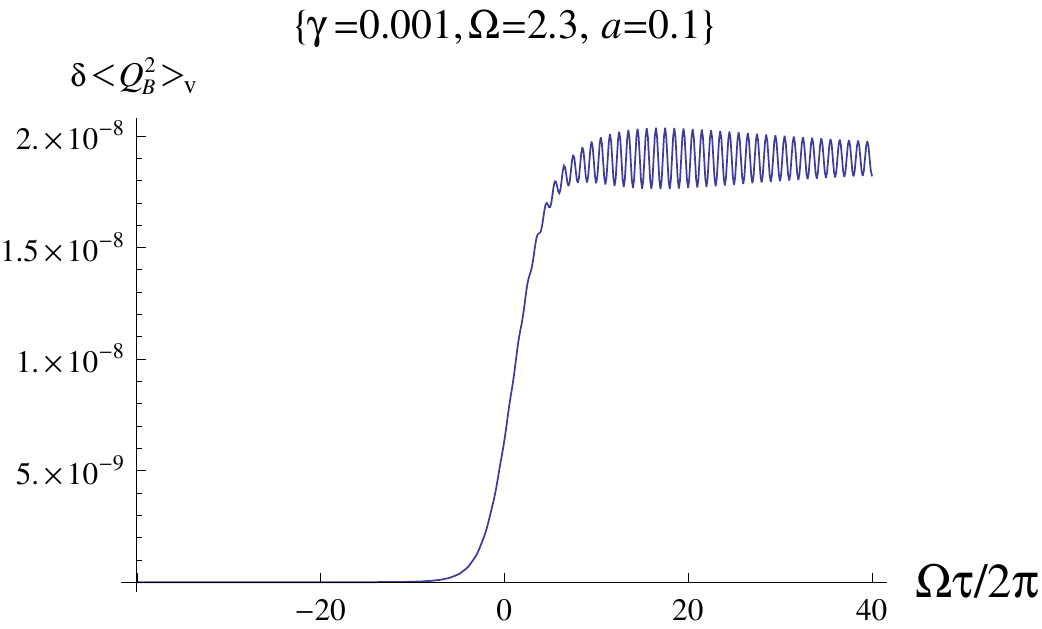}
\includegraphics[width=4.9cm]{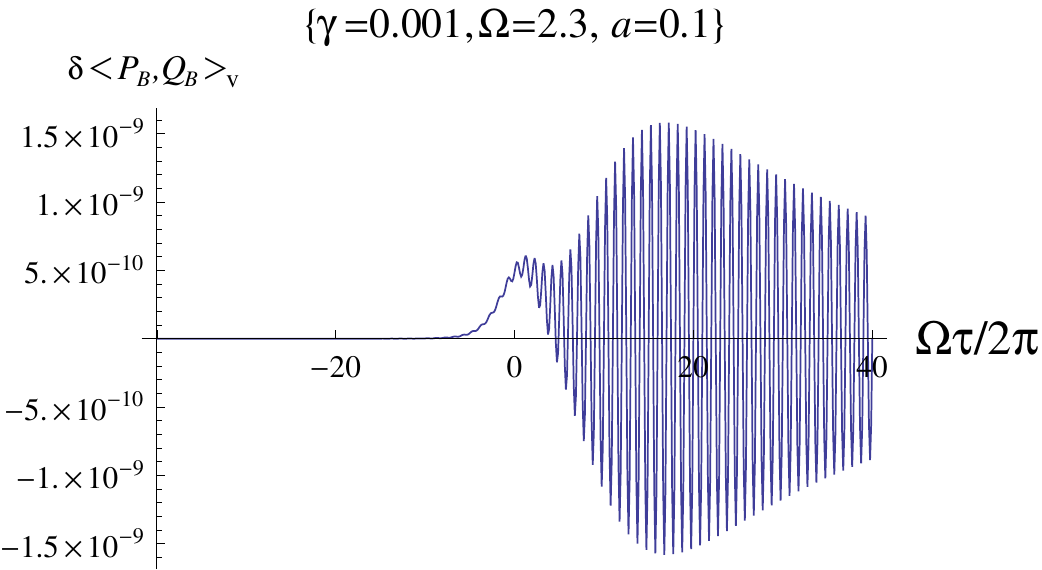}
\includegraphics[width=4.9cm]{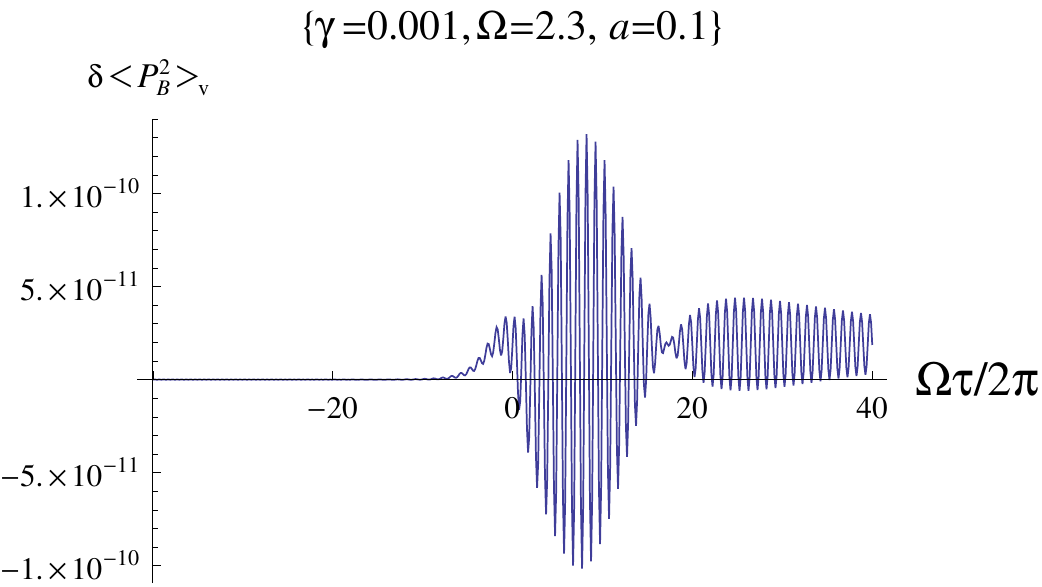}
\caption{Evolution of $\delta\left<\right. Q_B^2(\tau)\left.
\right>_{\rm v}$ (left), $\delta\left<\right. P_B,Q_B(\tau)\left.\right>_{\rm v}$
(middle), and $\delta\left<\right. P_B^2(\tau)\left.\right>_{\rm v}$
(right) in proper time $\tau$ of detector $B$ with a small $a$ and a weaker $\gamma$.
Here $\gamma=0.001$, $\Omega=2.3$, $m_0=\hbar=1$, and $a=0.1<1$.
The oscillations here are mainly produced by the impulse at the initial moment
and can be suppressed by choosing a more negative initial moment or a smaller $a$.}
\label{weakag}
\end{figure}

\begin{figure}
\includegraphics[width=4.9cm]{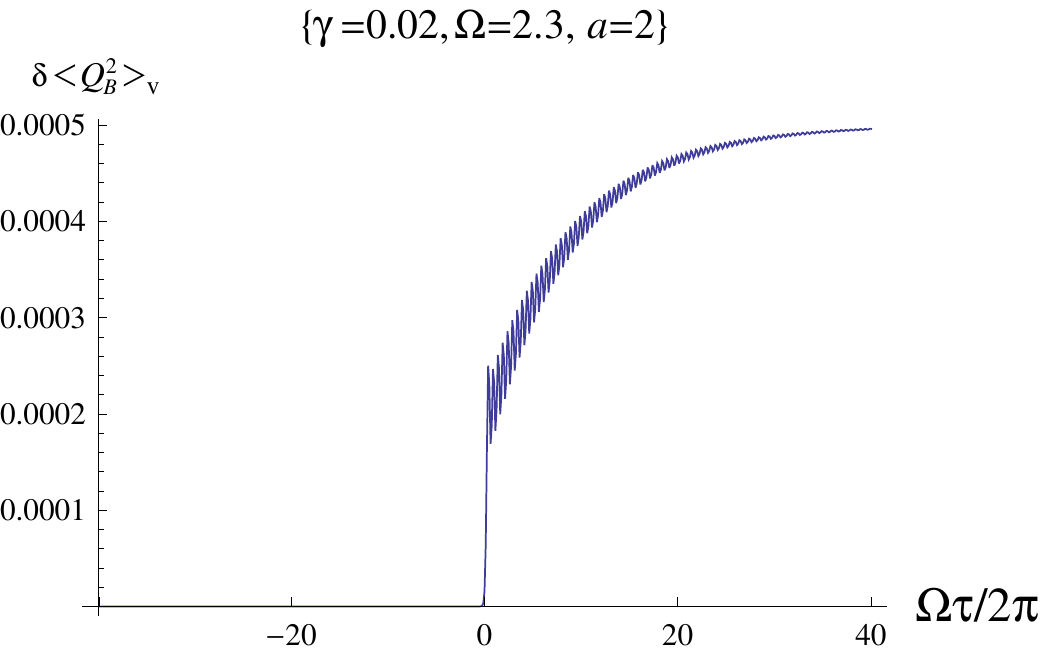}
\includegraphics[width=4.9cm]{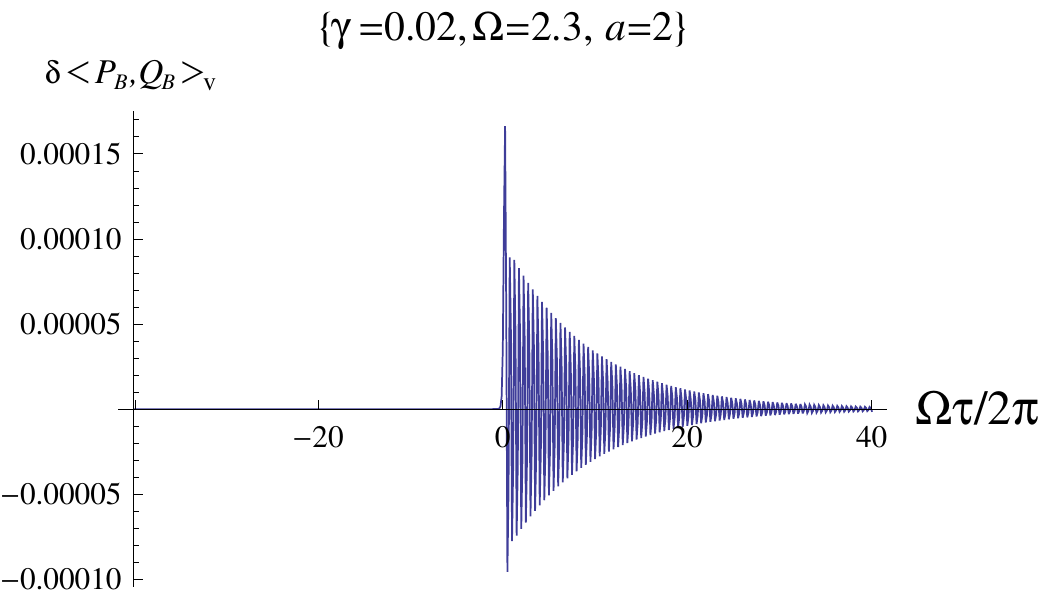}
\includegraphics[width=4.9cm]{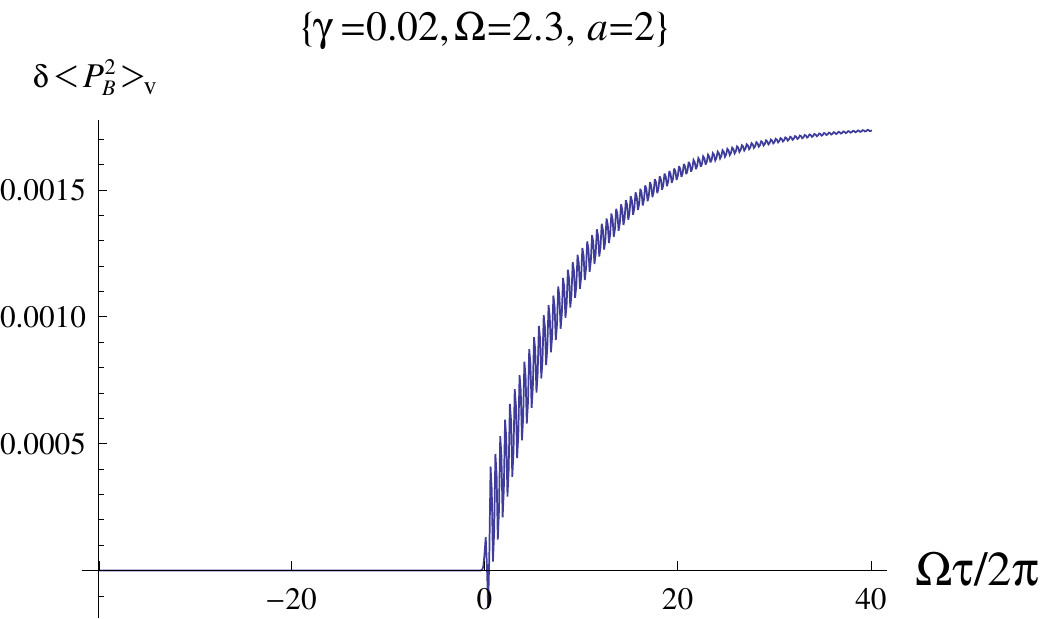}
\caption{Evolution of $\delta\left<\right. Q_B^2(\tau)\left.\right>_{\rm v}$
(left), $\delta\left<\right. P_B(\tau),Q_B(\tau)\left.\right>_{\rm v}$
(middle), and $\delta\left<\right. P_B^2(\tau)\left.\right>_{\rm v}$
(right) in proper time $\tau$ of detector $B$ with a large late-time proper
acceleration $a$. Here $\gamma=0.02$, $\Omega=2.3$, $m_0=\hbar=1$, and $a=2 >1$.
The oscillations here are due to a non-adiabatic effect during the 
transition and cannot be suppressed by choosing a more negative initial moment.}
\label{alarge}
\end{figure}

In those cases with small $a$  ($a < \Omega/\pi$ here),
the oscillations on top of the growth curves of $\delta\left<\right.Q_B^2(\tau)
\left.\right>_{\rm v}$ are mainly due to the impulse at the initial moment
(see Figs.~\ref{dQB2} (upper-left) and~\ref{weakag} (left)). One can see this by observing
that~(\ref{avary}) implies the proper acceleration
$\alpha^{}_B(\tau)\approx a e^{-3a|\tau|}$ when $a\tau\ll -1$ (as shown in the
upper-right plot of figure~\ref{dQB2}). Indeed, these oscillations 
(in the cases with $a < 2.3/\pi \approx 0.73$)
will be reduced if we choose a more negative initial moment $\tau_0$ 
or a larger $a$ (figure~\ref{dQB2} (upper-left)) to suppress the initial value of
the proper acceleration $\alpha^{}_B(\tau_0)$. However, in those cases with
larger late-time proper accelerations  ($a > \Omega/\pi$), though the impulse at the
initial moment is more suppressed, the amplitudes of those oscillations after
$\tau \approx 0$ become even larger  
but almost independent of the initial moment $\tau_0$ for $\Omega\tau_0\ll -1$, as illustrated 
in our numerical results in figure~\ref{dQB2} (lower-left) and figure~\ref{alarge} (left).
This indicates that these oscillations are coming from the non-adiabatic growth of 
the proper acceleration around $\tau = 0$ rather than the initial impulse.

To see the non-adiabatic behavior more closely, we set
 $\chi \equiv a \xi(\tilde{\tau})$ and $\chi' \equiv a \xi(\tilde{\tau}')$,
and rewrite the $\tilde{\tau}$-integrals into the $\chi$-integrals. 
eq.~(\ref{dQB2def}) then becomes
\begin{eqnarray}
  & & \delta\left<\right.Q_B^2(\tau)\left.\right>_{\rm v} = {2\gamma\hbar\over\pi m_0 \Omega^2}
  \times\nonumber\\ & & \hspace{.5cm}
  \int_{a\xi(\tau_0)}^{a\xi(\tau)} d\chi \int_{a\xi(\tau_0)}^{a\xi(\tau)} d\chi'
  K(\tau - \tau(\chi/a)) K(\tau - \tau(\chi'/a))  f(\chi,\chi'),
  \label{dQB2chi}
\end{eqnarray}
where
\begin{eqnarray}
f(\chi, \chi') &\equiv& \sqrt{e^{-2\chi}+1} \sqrt{e^{-2\chi'}+1} \times\nonumber\\& & \left\{
    {-1\over 4\left( 1 + e^{-(\chi+\chi')}\right)\sinh^2 \left({\chi- \chi'\over 2}\right)} +
     {1\over a^2\left[\tau(\chi/a)-\tau(\chi'/a) \right]^2} \right\}.
\label{fdef}
\end{eqnarray}
Here $\tau(\xi)$ was given in~(\ref{TauOfXi}),  implying that $f$ is
actually independent of $a$ in terms of $\chi$ and $\chi'$.
Now all the dependence on $a$ in $(\ref{dQB2chi})$ is coming from the $K$ functions
as well as the upper and the lower limits of the integration. In appendix \ref{jump} we can see
that the length scale of the non-trivial structure of the function $f$ in $\chi\chi'$-space is 
roughly of order $1$ (see Figure \ref{fcontour}), while the length scale of oscillations of $K$
in $\chi$ is about $a\pi/\Omega$ for $\chi > 0$. 
Thus for $a \ll \Omega/\pi$, $K$ oscillates so rapidly that the structure of $f$ will be
averaged out after integration and so $\delta\left<\right.Q_B^2(\tau)\left.\right>_{\rm v}$
evolves smoothly,  whereas for $a>\Omega/\pi$, the structure of $f$ could induce significant
non-adiabatic oscillations of $\delta\left<\right.Q_B^2(\tau)\left.\right>_{\rm v}$.

Nevertheless, since (\ref{dQB2def}) and the counterparts for other self correlators of detector $B$ are $O(\gamma)$,
the ``jumps" (\ref{dQBjump}) as well as the oscillations here are always small compared with the value of the 
self correlators themselves, which are $O(1)$  in the weak coupling limit. More details can be found in appendix \ref{jump}.

Similar quantities for the v-parts of other self correlators of detector $B$ 
can be obtained by replacing $K(x)$ by $K'(x) = dK(x)/dx$ whenever $Q_B(\tau(t))$ is 
replaced by $P_B(\tau(t))$ in~(\ref{dQB2def}). 
Two examples are shown in Figs.~\ref{weakag} and~\ref{alarge} . One can see that
$\delta\left<\right.P_B^2(\tau)\left.\right>_{\rm v}$ behaves roughly similar
to $\delta\left<\right.Q_B^2(\tau)\left.\right>_{\rm v}$, except for the larger
oscillations and the lack of a significant jump around the transition time
$\tau\approx 0$, while $\delta\left<\right.Q_B(\tau),
P_B(\tau)\left.\right>_{\rm v}$ is manifest only after $\tau\approx 0$.

\subsection{Cross correlators in Minkowski time}
\label{XcorrMt}

\begin{figure}
\includegraphics[width=6cm]{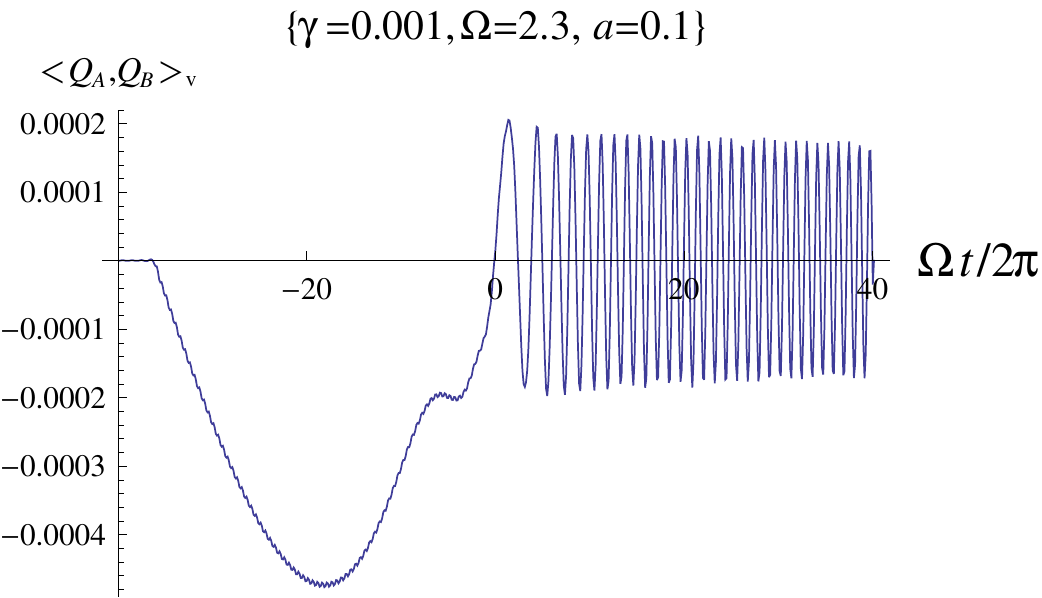}
\caption{Numerical results for the cross correlator
$\left<\right. Q_A(t), Q_B(\tau(t)) \left.\right>_{\rm v}$ with $d=10$
evolving in Minkowski time $t$. Other cross correlators $\left<\right.
Q_A,P_B\left.\right>_{\rm v}$, $\left<\right. P_A,Q_B\left.\right>_{\rm v}$,
and $\left<\right. P_A, P_B\left.\right>_{\rm v}$ have similar behaviors.}
\label{XcorrEx}
\end{figure}

From~(\ref{zA}) and~(\ref{zB}), we have
\begin{eqnarray}
  && \left<\right. Q_A(t), Q_B(\tau(t))\left.\right>_{\rm v}^{(0)}\nonumber\\
  &=& {2\gamma\hbar\over\pi m_0\Omega^2}
    {\rm Re} \int_{t_0}^t d\tilde{\tau} \int_{\tau(t_0)}^{\tau(t)} d\tilde{\tau}'
    \times\nonumber\\& &\hspace{.5cm}
    {K(t-\tilde{\tau})K(\tau(t)-\tilde{\tau}')\over \left[ -d -
    {1\over a}\cosh a\tilde{\xi}' +{1\over 2a}e^{-a\tilde{\xi}'}\right]^2
    -\left[ \tilde{\tau}-{1\over a}\sinh a\tilde{\xi}' +
    {1\over 2a}e^{-a\tilde{\xi}'} -i\epsilon\right]^2 } \nonumber\\
  &=& {2\gamma\hbar\over\pi m_0\Omega^2} {\rm Re} \int_{t_0}^t d\tilde{\tau}
    \int_{\tau(t_0)}^{\tau(t)} d\tilde{\tau}' {K(t-\tilde{\tau})K(\tau(t)-\tilde{\tau}')
    \over 2d + a^{-1} e^{a\tilde{\xi}'}}\times\nonumber\\& &\hspace{.5cm}
    \left[{1\over\tilde{\tau}+d+a^{-1}e^{-a\tilde{\xi}'}
    -i\epsilon} -{1\over \tilde{\tau}-d- 2 a^{-1}\cosh a\tilde{\xi}'-i \epsilon}\right],
\label{XcorrInt}
\end{eqnarray}
with $\tilde{\xi}' \equiv \xi(\tilde{\tau}'+i\epsilon)$. 
The v-parts of other cross correlators can be obtained by replacing $K(t-\tilde{\tau})$ by
$K'(t-\tilde{\tau})$ whenever $Q_A(t)$ is replaced by $P_A(t)$, and replacing
$K(\tau(t)-\tilde{\tau}')$ by $K'(\tau(t)-\tilde{\tau}')$ whenever
$Q_B(\tau(t))$ is replaced by $P_B(\tau(t))$. An example of the cross
correlators is shown in figure~\ref{XcorrEx}. One can see that the early-time
behavior of the cross correlators in this setup is quite similar to those of
two inertial detectors~\cite{LH09} (see also figure~\ref{Xcomp}). 
The absolute values of the cross correlators start to grow significantly after 
each detector enters the other's light cone, then keep growing until the motion 
of detector $B$ becomes obvious. After $t\approx 0$, the behavior of the cross 
correlators turns to a fashion similar to those in the case of uniformly accelerated 
detectors~\cite{LCH08, LH10}. They become oscillating in $t$ with amplitude decaying as
$e^{-2\gamma t}$.

Comparing the results in figure~\ref{XcorrEx} and figure~\ref{weakag} with the same
parameters ($\gamma=0.001$, $\Omega=2.3$, $a=0.1$), we find that even for
$d=10$, which is not very small, the values of the v-part of the cross
correlators in this parameter regime are much greater than  those of the
deviations of the v-part of the self correlators from their zero-acceleration
limits.

\subsection{Mutual influences}
\label{secmuinf}

From the equations of motion for the mode functions eqs.~$(13)$--$(16)$
in~\cite{LCH08}, the mode functions with corrections from mutual influences can
be written as
\begin{equation}
  q^{(\mu)}_j = q^{(\mu)(0)}_j + \sum_{n=1}^{\infty} q^{(\mu)(n)}_j,
\label{qhigh}
\end{equation}
where $q^{(\mu)(0)}_j$ are the zeroth order  solutions
without considering mutual influences, and
\begin{equation}
  q^{(\mu)(n)}_j(\tau) = {2\gamma\over\Omega} \int_{\tau_0}^\tau d\tilde{\tau}
    \theta\left(\tau_{\bar{j}}^{ret}(\tilde{\tau}) - \tau_{\bar{j}}^{(n-1)}\right)
    K(\tau-\tilde{\tau}){q^{(\mu)(n-1)}_{\bar{j}}(\tau^{ret}_{\bar{j}}(\tilde{\tau}))
    \over R_{\bar{j}\to j}(\tilde{\tau})}
\label{qmun}
\end{equation}
with $\mu \in \{A,B,+,-\}$, $i,j\in\{A, B\}$, $\bar{A}\equiv B$, $\bar{B}\equiv
A$, the retarded times $\tau_A^{ret}(\tilde{\tau})=
t^{ret}(z^\mu_B(\tilde{\tau}))$ and $\tau_B^{ret}(\tilde{\tau}) =
\tau(\xi^{ret}(z^\mu_A(\tilde{\tau})))$, $\tau_j^{(n)} \equiv
[\tau_j^{ret}]^{-1}(\tau_{\bar{j}}^{(n-1)})$, $\tau_A^{(0)} \equiv t_0$,
$\tau_B^{(0)} \equiv \tau(\xi(t_0))$, and the retarded distance $R_{\bar{j}\to
j}$ defined in appendix~\ref{retard} (see figure~\ref{defmutinf}).

For the a-part of the correlators, it is straightforward to obtain the
corrected results by simply inserting the corresponding~(\ref{qhigh})
into eq.~$(25)$ in~\cite{LCH08}. For the v-part of the correlators, the calculation
is not as straightforward because of the mode sum $\int d^3 k$. Nevertheless,
we can express the corrected correlators up to the $N$-th order mutual
influences as, for example,
\begin{equation}
  \left< \right. Q_i(\tau_i), Q_j(\tau_j)\left.\right>_{\rm v, a} \approx
  \sum_{m=0}^N \sum_{n=0}^N \left< \right. Q_i^{(m)}(\tau_i), Q_j^{(n)}(\tau_j)\left.\right>_{\rm v, a},
\end{equation}
where the $(m,n)$-th order correlator $\left< \right. Q_i^{(m)}(\tau_i),
Q_j^{(n)}(\tau_j) \left.\right>$ can be obtained recursively from those of
lower orders by
\begin{eqnarray}
  & &\left< \right. Q_i^{(m)}(\tau_i), Q_j^{(n)}(\tau_j)\left.\right>_{\rm v, a} = 
    {2\gamma\over\Omega}  \times \nonumber\\ & & \hspace{.5cm}
    \int_{\tau_j^{(0)}}^{\tau_j}d\tilde{\tau}\,
    \theta\left(\tau_{\bar{j}}^{ret}(\tilde{\tau}) - \tau_{\bar{j}}^{(n-1)}\right)
    {K(\tau_j-\tilde{\tau})\over R_{\bar{j}\to j}(\tilde{\tau})}
    \left< \right. Q_i^{(m)}(\tau_i), Q_{\bar{j}}^{(n-1)}(\tau_j^{ret}(\tilde{\tau}))
    \left.\right>_{\rm v, a}
  \label{vhighmn}
\end{eqnarray}
($m,n \ge 1$) and their $\tau_i$ or $\tau_j$ derivatives.

\begin{figure}
\includegraphics[width=6cm]{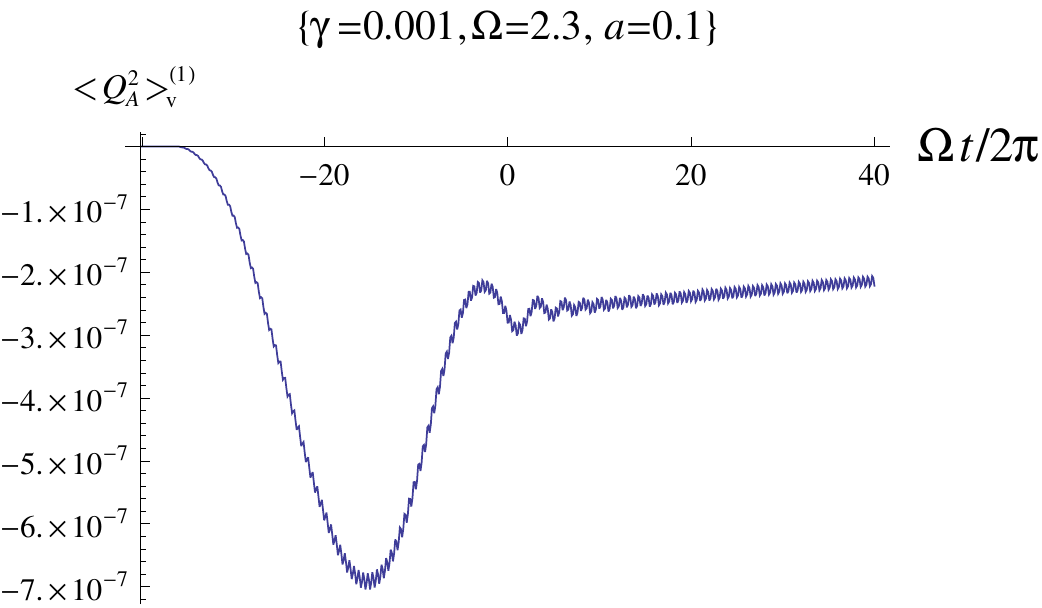}
\includegraphics[width=6cm]{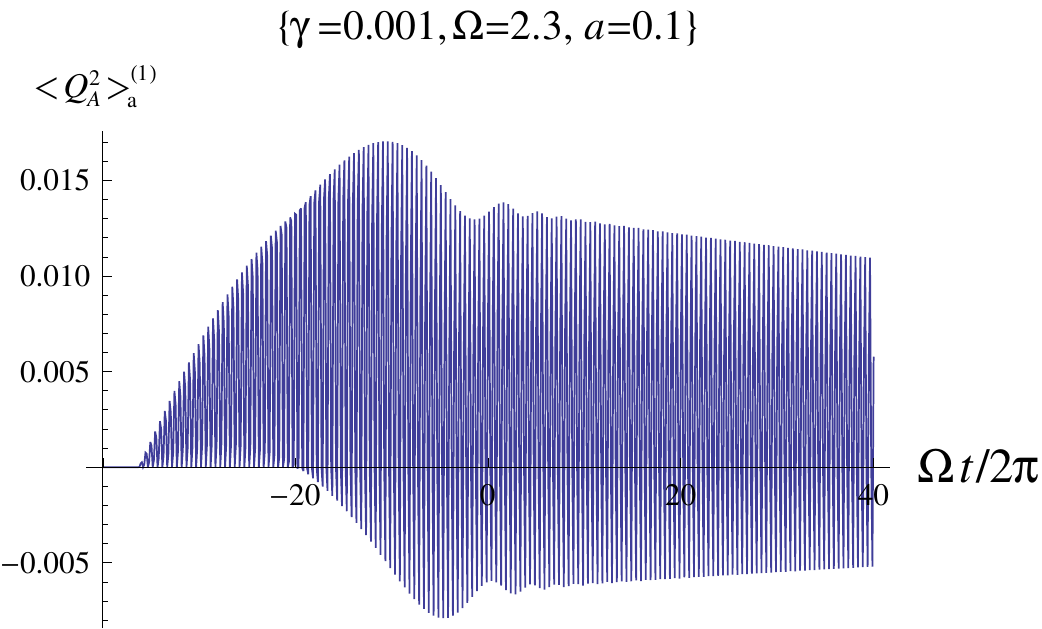}
\caption{Numerical results for the first order corrections to $\left< \right. Q_A^2(t)
\left.\right>_{\rm v}$ (left) and $\left< \right. Q_A^2(t)\left.\right>_{\rm a}$ (right)
from mutual influences with $d=10$, $\alpha=1.4$, and $\beta=0.2$. One can see the
profile of the envelope of the oscillating $\left< \right. Q_A, Q_B\left.\right>_{\rm v}$
in figure~\ref{XcorrEx}.} 
\label{muinf1}
\end{figure}

From~(\ref{qhigh})--(\ref{vhighmn}) we see that the $N$-th order
corrections are roughly $O((\gamma/\Omega d)^N)$ 
compared with the magnitude of the zeroth order correlators.
The presence of the oscillating function $K(\tau-\tilde{\tau})$ in the integrand of~(\ref{qmun})
further indicates that one detector ($j$) will be influenced very little by the off-resonant part of
$q^{(\mu)(n-1)}_{\bar{j}}(\tau^{ret}_{\bar{j}}(\tilde{\tau}))/R_{\bar{j}\to j}
(\tilde{\tau})$ from the other detector ($\bar{j}$) in the weak coupling limit.
In our setup since the trajectories
of the two detectors are asymmetric, the retarded field solution from detector $j$ will
always be red- or blue-shifted in view of detector $\bar{j}$. So most of
$q^{(\mu)(n-1)}_{\bar{j}}(\tau^{ret}_{\bar{j}}(\tilde{\tau}))/R_{\bar{j}\to j}(\tilde{\tau})$
are off-resonant and thus mutual influences can be very small even though the magnitudes of
$q^{(\mu)(n-1)}_{\bar{j}}(\tau^{ret}_{\bar{j}}(\tilde{\tau}))/R_{\bar{j}\to j}(\tilde{\tau})$
appear larger.

According to our analysis and numerical results, mutual influences are indeed negligible
in perturbative regime when the distance between the detectors is always large,
i.e., $\gamma/\Omega d \ll 1$. For example, our numerical results
show that when $\gamma=10^{-3}$, $\Omega =2.3$,
$a=0.1$, $d=10$, $\Lambda_0=\Lambda_1 = 20$, the magnitudes of the first order corrections
$\left< \right. Q_A^2(t)\left.\right>_{\rm v}^{(1)} \equiv
2\left< \right. Q_A^{(0)}(t), Q_A^{(1)}(t)\left.\right>_{\rm v} +
\left< \right. Q_A^{(1)}(t), Q_A^{(1)}(t)\left.\right>_{\rm v}$ are less than $10^{-4}$ of
the magnitude of $\left< \right. Q_A^{(0)}(t), Q_A^{(0)}(t)\left.\right>_{\rm v}$
(see Figs.~\ref{muinf1} (left) and~\ref{corrv} (left)),
while the magnitudes of the first order correction
$\left< \right. Q_A^2(t)\left.\right>_{\rm a}^{(1)} \equiv
2\left< \right. Q_A^{(0)}(t), Q_A^{(1)}(t)\left.\right>_{\rm a} +
\left< \right. Q_A^{(1)}(t), Q_A^{(1)}(t)\left.\right>_{\rm a}$ are about $10^{-3}$ that of
$\left< \right. Q_A^{(0)}(t), Q_A^{(0)}(t)\left.\right>_{\rm a}$
(see Figs.~\ref{muinf1} (right) and~\ref{corra} (left)).
The corrections   become even smaller as we decrease $\gamma$ or increase $d$.
The ratio $\left< \right. Q_A^2(t)\left.\right>_{\rm v}^{(1)}/
\left< \right. Q_A^2(t)\left.\right>_{\rm v}^{(0)}$ is smaller than $\left< \right. Q_A^2(t)
\left.\right>_{\rm a}^{(1)}/\left< \right. Q_A^2(t)\left.\right>_{\rm a}^{(0)}$ simply because
the v-part of the zeroth order cross correlator $\left< \right. Q_A(t),Q_B(t)\left.\right>_{\rm v}^{(0)}$
is suppressed when $d$ is large (see eq.~(\ref{Xcorrd})), while
$\left< \right. Q_A(t),Q_B(t)\left.\right>_{\rm a}^{(0)}$ is independent of $d$.

\subsection{Entanglement dynamics with weak coupling and large separation}
\label{EntDyn}

\begin{figure}
\includegraphics[width=4.9cm]{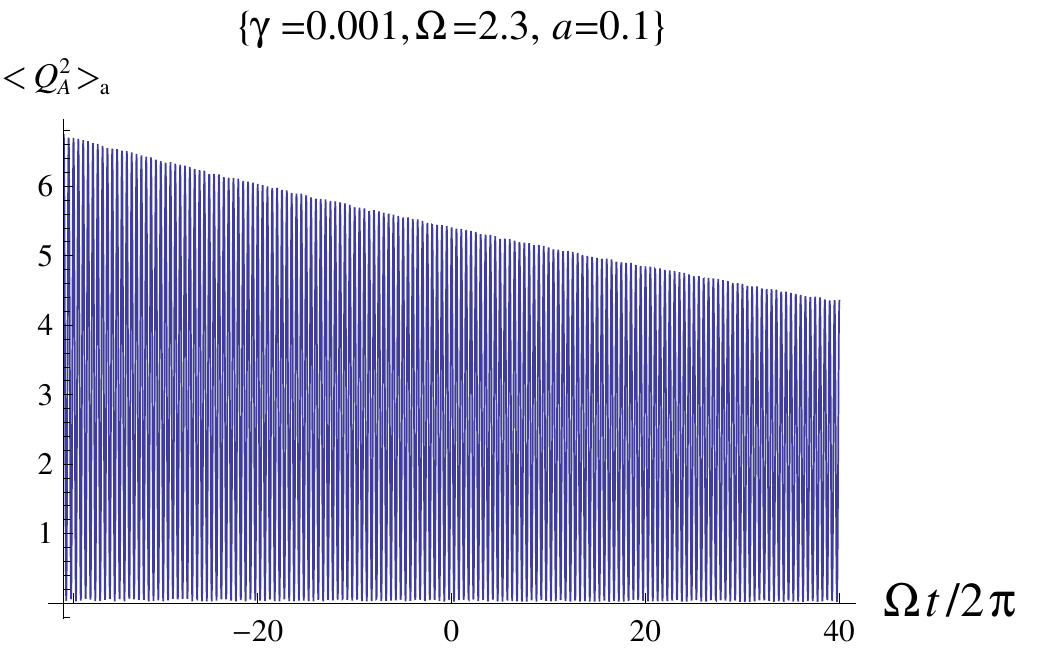}
\includegraphics[width=4.9cm]{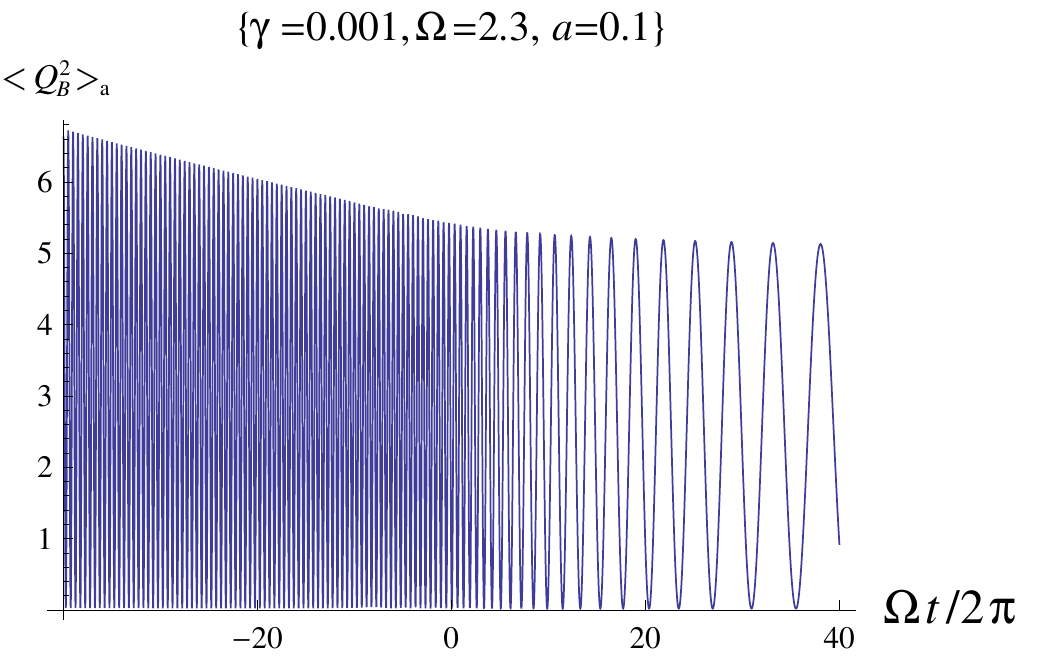}
\includegraphics[width=4.9cm]{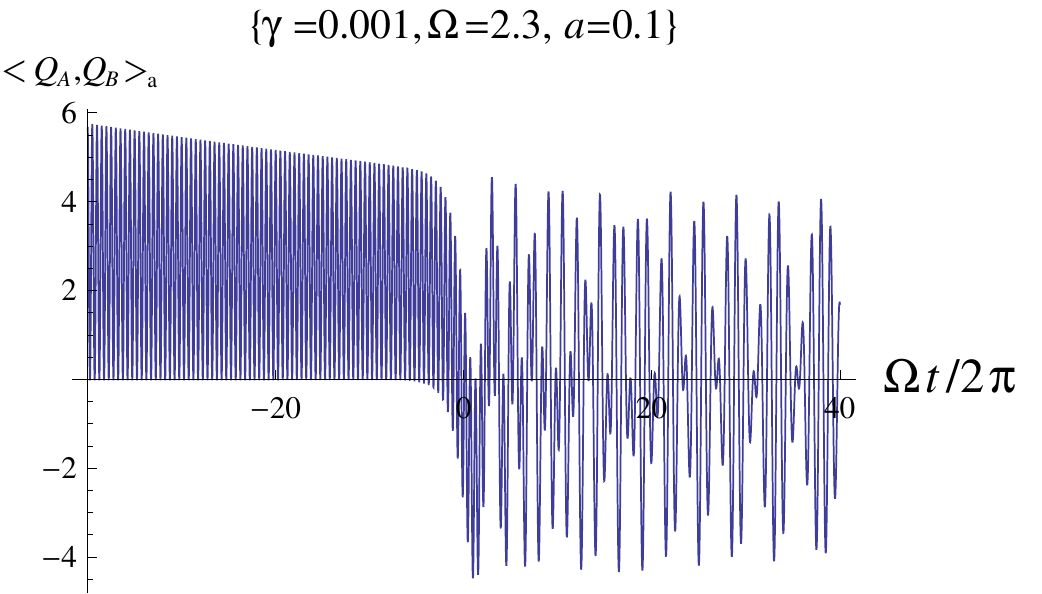}
\caption{The evolution of
$\left< \right. Q_A^2 \left.\right>_{\rm a}$ (left),
$\left< \right. Q_B^2 \left.\right>_{\rm a}$ (middle), and
$\left< \right. Q_A, Q_B \left.\right>_{\rm a}$ (right)
with $\alpha=1.4$, $\beta=0.2$ and $d=10$ in Minkowski time.
The oscillations are consequences of choosing the initial state of the detectors
as a squeezed state. Time dilation in the results for detector $B$ is manifest
after $at > -0.114$, which makes the cross correlators behave more irregularly.}
\label{corra}
\end{figure}

\begin{figure}
\includegraphics[width=6cm]{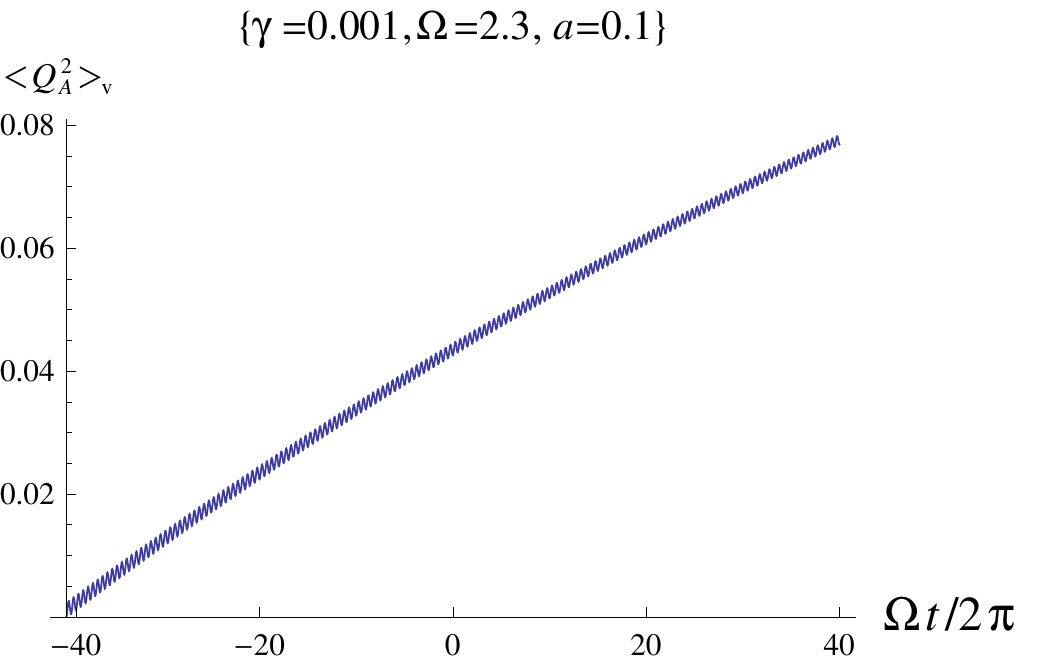}
\includegraphics[width=6cm]{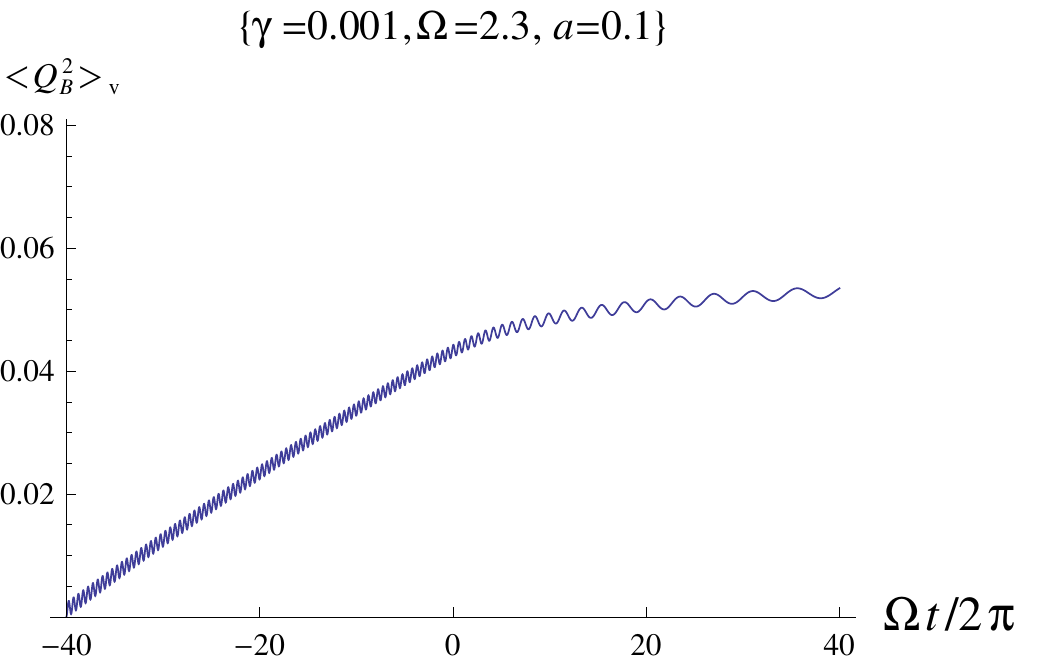}
\caption{Evolution of the v-part of the self correlators $\left<\right.  Q_A^2 \left.\right>_{\rm v}$
and $\left<\right. Q_B^2 \left.\right>_{\rm v}$ in Minkowski time with
$d=10$ and $\Lambda_0=\Lambda_1=20$. Again, time dilation in the results for
detector $B$ is manifest after $at > -0.114$. The contributions from the
differences shown in figure~$\ref{weakag}$ are very small and not significant in
these plots.}
\label{corrv}
\end{figure}

Combining all the above elements with weak coupling and large separation,
examples of the evolution of the a-part of
the correlators are shown in figure~\ref{corra}, while those of the v-part of the
self correlators are shown in figure~\ref{corrv}.

The dynamics of quantum entanglement between the two detectors in Gaussian
state can be found straightforwardly by examining the behavior of the quantity
$\Sigma$~\cite{LCH08, Si00} and the logarithmic negativity $E_{\cal N}$~\cite{VW02}
defined by
\begin{eqnarray}
  \Sigma &\equiv&\det\left[ {\bf V}^{PT}+{i\hbar\over 2}{\bf M}\right]
    =\left( c_+^2 -{\hbar^2\over 4}\right)\left( c_-^2 -{\hbar^2\over 4}\right),\\
  E_{\cal N} &\equiv& \max \left\{ 0, -\log_2 2c_- \right\},
\end{eqnarray}
where ${\bf M}$ is the symplectic matrix ${\bf 1}\otimes (-i)\sigma_y$,
${\bf V}^{PT}$ is the partial transpose $(Q_A, P_A, Q_B, P_B)$ $\to
(Q_A, P_A, Q_B, -P_B)$ of the covariance matrix ${\bf V}$ in $(\ref{coVm})$, and
$(c_+, c_-)$ is the symplectic spectrum of ${\bf V}^{PT}+ (i\hbar/2){\bf M}$, given by
\begin{equation}
  c_\pm \equiv \left[Z \pm \sqrt{Z^2-4\det {\bf V}}\over 2\right]^{1/2}
\label{SympSpec}
\end{equation}
with $Z \equiv \det {\bf v}^{}_{AA} + \det {\bf v}^{}_{BB} - 2 \det {\bf v}^{}_{AB}$.
For the detectors in a Gaussian state, the reduced state of the detectors is
entangled if and only if $c_- < \hbar/2$~\cite{Si00}, when $E_{\cal N}>0$ and
$\Sigma <0$. The value of $E_{\cal N}$ indicates the degree of entanglement: we
say the two detectors have a stronger entanglement if the associated
$E_{\cal N}$ is greater.  However it is more convenient to use $\Sigma$ in
calculating the disentanglement time~\cite{LCH08, LH09}.

\begin{figure}
\includegraphics[width=4.9cm]{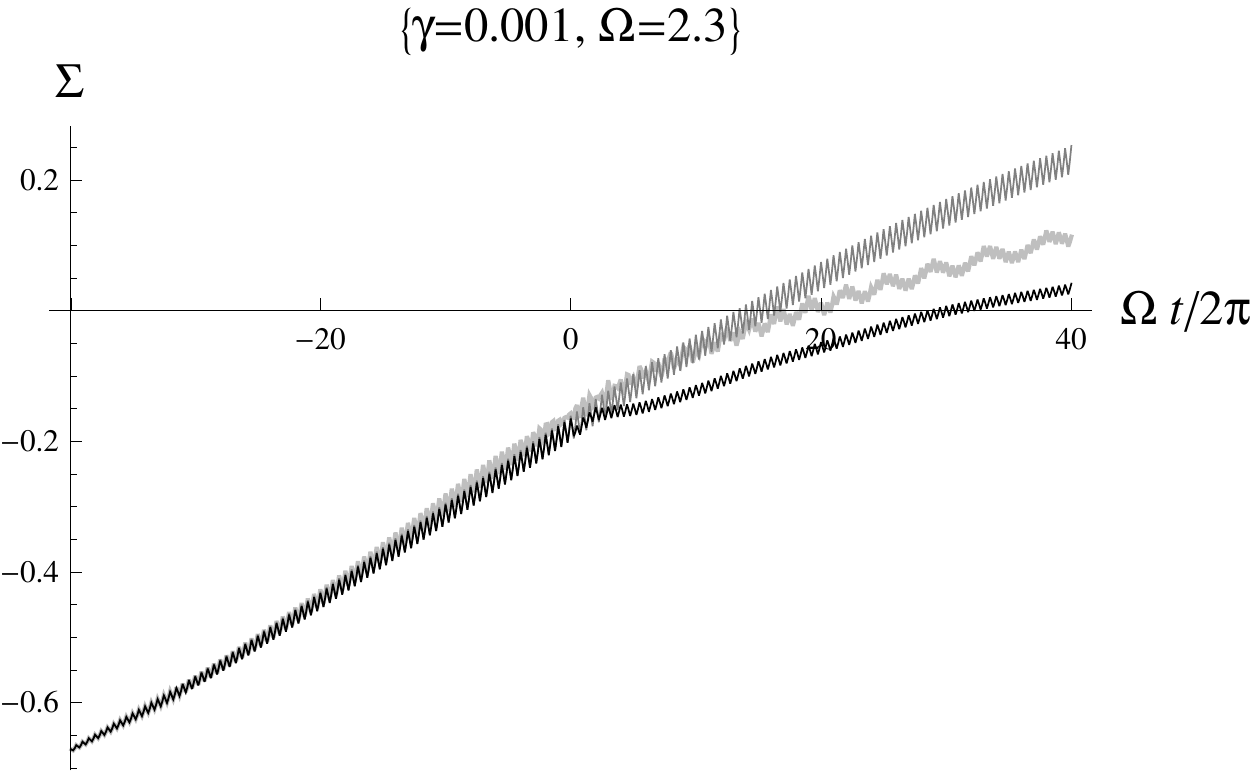}
\includegraphics[width=4.9cm]{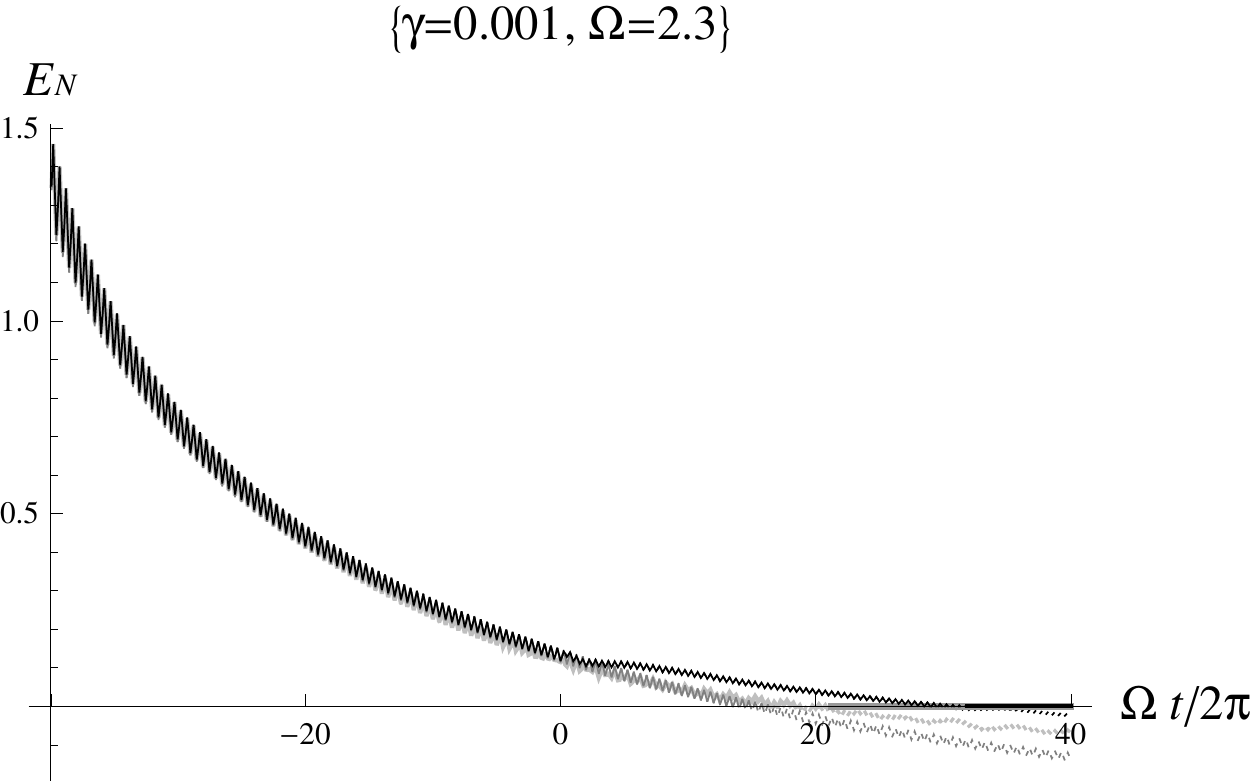}
\includegraphics[width=4.9cm]{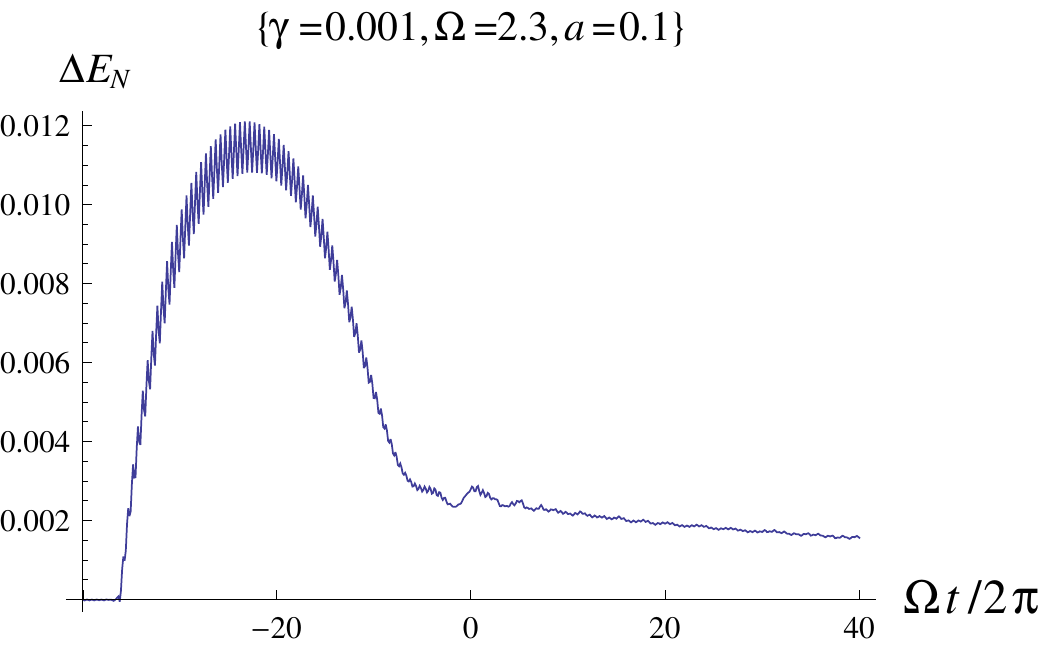}
\caption{
Numerical results with $\alpha=1.4$, $\beta=0.2$ in~(\ref{initGauss}) for the
quantity $\Sigma(t)$ (left) the logarithmic negativity $E_{\cal N}(t)$ (middle,
solid curves), both indicating the degree of entanglement between the detector
at~(\ref{zA}) with $d=10$ and the detector going along~(\ref{zB}). The gray,
the thick-lightgray, and the black curves in both plots represent the results
with $a=0$, $0.1$, and $2$, respectively, where the $a=0$ case corresponds to
those for two inertial detectors both at rest in space and separated at a
distance $d=10$ in the same initial state~\cite{LH09}. Quantum entanglement
experiences sudden death at $(\Omega t_{dE}/2\pi) \approx 16$ for $a=0.1$ and
$\approx 30$ for $a=2$ when $E_{\cal N}$ touches $0$ and $\Sigma$ crosses $0$.
One can see that the larger the value of $a$, the longer the disentanglement
time $t_{dE}$, due to the time dilation of the  moving detector $B$ in this setup.
(Right) The difference
$\Delta E_{\cal N}\equiv E_{\cal N} - (E_{\cal N}|_{\left<\right.
{\cal R}_A(t), {\cal R'}_B(\tau(t)) \left.\right>_{\rm v}=0})$
for $a=0.1$. One can see the profile of the envelopes of the oscillating
cross correlators in figure~\ref{XcorrEx}. The value of the deference is tiny
compared with the value of $E_{\cal N}$ (the largest ratio is about $3\%$
around $t\approx -18 (2\pi)/\Omega$).}
\label{entdyn}
\end{figure}

An example of the sudden death of entanglement is given in figure~\ref{entdyn},
where we see that the curves  with $a\not=0$ are stretched horizontally after
$t\approx 0$ due to the time dilation of the moving detector $B$. This
increases the disentanglement time.

The contribution by the v-part of the cross correlators to entanglement
dynamics is suppressed efficiently when the coupling is weak
(here $\gamma=0.001$, not quite weak, though) and the distance is large (here
$d=10$). In the difference
$E_{\cal N} -(E_{\cal N}|_{\left<\right. {\cal R}_A(t), {\cal R}_B(\tau(t)) \left.\right>_{\rm v}=0})$
shown in figure~\ref{entdyn} (middle), we recognize the profile of the envelopes
of the oscillating cross correlators in figure~\ref{XcorrEx}. Thus the nonvanishing
cross correlators tend to enhance the degree of entanglement between the
detectors. However the enhancement of $E_{\cal N}$ is tiny --- so tiny that it
is safe to neglect the v-part of the cross correlators and skip the
time-consuming computation for them in the weak-coupling limit with long
initial distances and large initial entanglement between the detectors.

In contrast, while the value of each v-part of the self correlators is small
compared with its a-counterpart, they are crucial in obtaining the entanglement
dynamics. If one sets all the v-parts of the self correlators to zero, the
evolution of $\Sigma$ and the logarithmic negativity $E_{\cal N}$ will be very
different (for example, see~\cite{LCH08}.) Therefore in the perturbative regime
with large initial distance  and entanglement  between the detectors, the
zeroth order of the a-part of all correlators as well as the v-part of the self
correlators are enough to give the entanglement dynamics to  high accuracy.

\section{Summary}
\label{summa}

We have demonstrated that the dynamics of a UD detector in non-uniform
acceleration are similar to those of a harmonic oscillator in contact with a
``thermal'' bath at a time-varying ``temperature''
in the weak coupling regime,
while non-adiabatic changes of proper acceleration will create oscillations on top of
the smoothly evolving values of the correlators. The behavior of the detector is
determined by the kinematics in its history 
rather than by assuming the presence of a horizon
that does not exist until late time.

In our model with weak coupling to the field, large spatial separation and
large initial entanglement between the detectors,  the higher-order corrections
from mutual influences are negligible and  the early-time behavior of the
detectors are dominated by the zeroth order of the a-parts of the self and
cross correlators of the detectors, which correspond to the initial state of
the detectors. The zeroth order contribution of the v-parts of the {\it self}
correlators of the detectors, which corresponds to the response of the
detectors to the field, is also crucial for entanglement dynamics, though their
values are small compared to their a-counterparts. While the zeroth order of
the v-part of the {\it cross} correlators would in general enhance quantum
entanglement between the detectors, their values are even smaller than others
and negligible in the perturbative regime if the initial degree of entanglement
between the detectors is large.

We have chosen a trajectory for  detector $B$ such that it is approximately at
rest when its proper time $\tau$ is negatively large, and almost uniformly
accelerated when $\tau$ is positively large. As  expected, the entanglement
dynamics of the detectors here are similar to those in the case of two inertial
detectors~\cite{LH09} when $\tau$ is negatively large, and look like those in
the case with one inertial detector and one uniformly accelerated detector when
$\tau$ is positively large~\cite{LCH08}. These results are commensurate with
those obtained previously using alternative methods for evaluating entanglement
dynamics of detectors in relative non-uniform acceleration~\cite{MV09}. 
Note that in \cite{PV92} Percocco and Villalba computed the Bogoluibov coefficients of a quantum 
field in the spacetime (\ref{CVframe}) and obtained a Planckian spectrum with exactly constant temperature 
parameter in the asymptotic limit. Nevertheless, it is not clear their temperature is well-defined 
since their time derivative $\partial_u$ is not a Killing vector.

While the dynamics of the correlators are more subtle during the transition of
detector $B$ from zero to finite accelerations, such interesting behavior is
negligible in computing the entanglement dynamics in the perturbative regime.
In our model we do see sudden death of entanglement (see figure~\ref{entdyn}). As
noted earlier in~\cite{LCH08}, however, the acceleration in this case increases
rather than decreases the disentanglement time because of the time dilation of
the moving detector $B$ observed in the conventional Minkowski coordinate,
though a higher Unruh temperature is experienced by detector $B$ at late times.

A number of interesting directions for further research emerge based on our
results. A time-reversed setup where detector $B$ begins in the distant past as
almost uniformly accelerated then becomes approximately inertial in the distant
future could be considered. Combining these results with those obtained in this
paper, the case with the world lines of the detectors similar to the ones in
the twin paradox~\cite{RHK} becomes straightforward in weak coupling limit with
large spatial separation~\cite{LinBehHu}. Extending our work to cosmological
settings that go beyond idealizations previously considered~\cite{cosmo} is
another avenue for further research.  Wider parameter ranges, such as those
beyond weak coupling, small acceleration and/or large spatial separation
regimes are also worth studying. 
By using the well-known correspondence between the Rindler and the Schwarzschild spacetimes, 
one can apply the knowledge obtained in this paper and go beyond the test-field description of 
black hole physics \cite{LCH08}. Regarding to the exchange of information, the setup in this paper
can also be applied to quantum teleportation between a free-falling agent and an initially free-falling agent 
who eventually stays outside the black hole \cite{LSCH12}.

\acknowledgments BLH and SYL wish to thank the hospitality of the
Perimeter Institute for hosting their visits in Spring 2008 where this joint
work began. This work is supported in part by the Natural Sciences and
Engineering Research Council of Canada, the NSF Grant No. PHY-0801368, 
the Nation Science Council of Taiwan under the Grant No. NSC 99-2112-M-018-001-MY3, 
and the National Center for Theoretical Sciences, Taiwan.                                                        

\begin{appendix}

\section{Retarded distance and retarded time}
\label{retard}

\begin{figure}
\includegraphics[width=5cm]{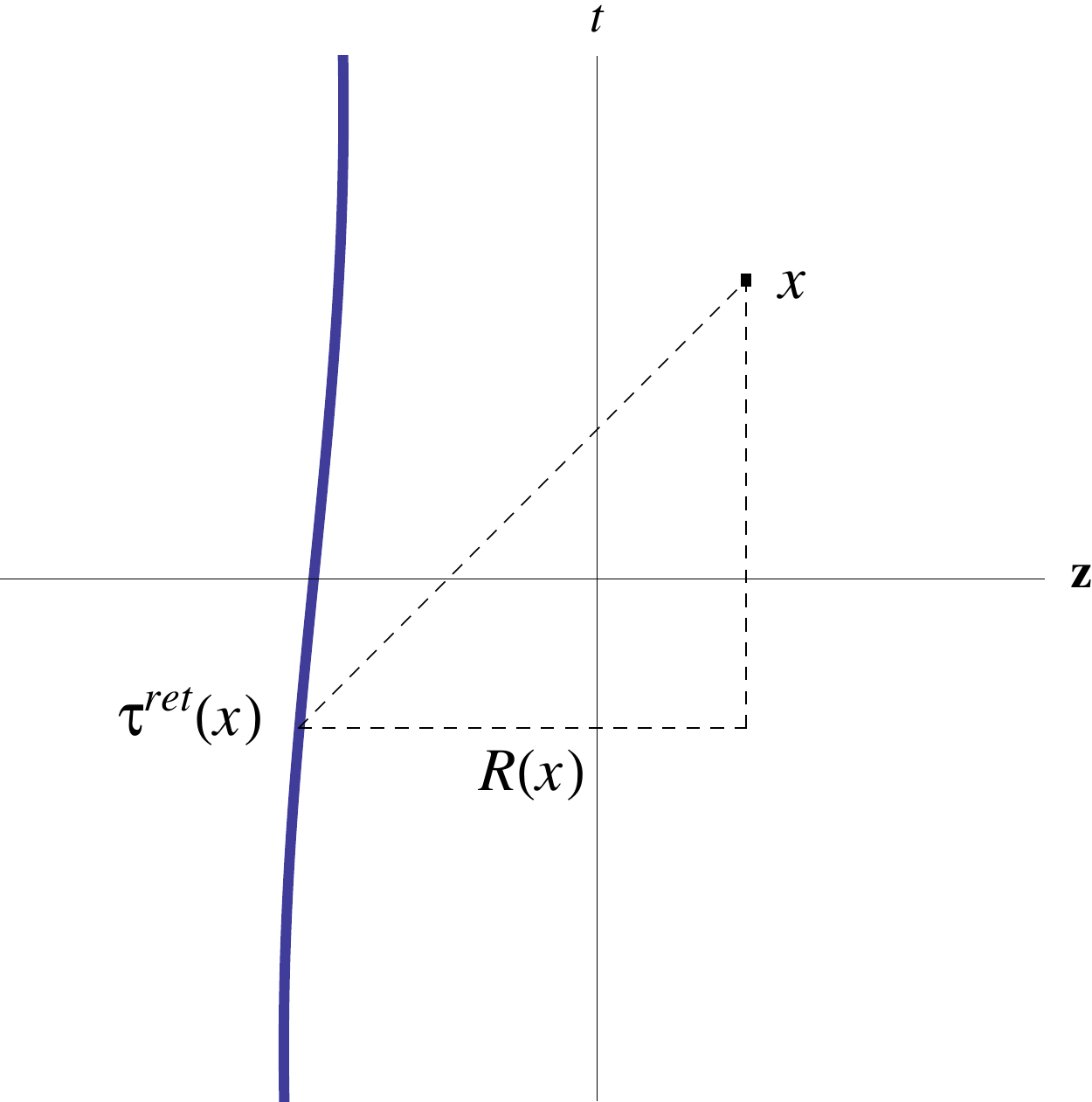}
\caption{Definitions of $\tau^{ret}(x)$ and $R(x)$. The thick curve is the
world line of a detector parametrized by its proper time $\tau$.}
\label{deftretR}
\end{figure}

For a massless field in (3+1)D flat spacetime, the retarded time
$\tau^{ret}(x^\mu)$ associated with a field observed at the spacetime point
$x^\mu$ is defined as the proper time $\tau$ of the field source at which the
trajectory of the point-like source $z^\mu(\tau)$ intercepts the past light
cone of $x^\mu$. It is given by the solution to
$\sigma(x^\mu,z^\mu(\tau^{ret})) =0$ where
\begin{equation}
  \sigma(x^\mu,z^\mu(\tau)) = -{1\over 2}\left( x^\mu - z^\mu(\tau)\right)
  \left( x_\mu - z_\mu(\tau)\right)
\end{equation}
is Synge's world function. Since $\sigma$ is quadratic, when $\sigma=0$ is
satisfied, the Dirac delta function $\delta(\sigma)$ in the retarded Green's
function of the field will give an $1/R$ factor in $\tau$-integrals involving
it, where
\begin{equation}
  R=\left| {d\sigma\over d\tau}\right|_{\sigma=0}
\end{equation}
is a function of $x^\mu$ and the location of the source at the retarded time
$\tau^{ret}(x^\mu)$. We call $R$ the retarded distance (see
figure~\ref{deftretR}).

For detector $A$, one has
\begin{equation}
  \sigma = -{1\over 2}\left[ D({\bf x})^2- (t-x^0)^2 \right],
\end{equation}
where $D({\bf x})\equiv \sqrt{(x^1 + d)^2+\rho^2}$ and $\rho^2\equiv x_2^2+ x_3^2$.
Solving $\sigma=0$, the retarded time of detector $A$ is found to be
\begin{equation}
  \tau_A^{ret}(x) = t^{ret}(x) \equiv x^0 - D({\bf x}).
\end{equation}
So the retarded distance is $R=D({\bf x})$. In particular, at the position of
the detector $B$, the retarded distance from $A$ to $B$ is
\begin{equation}
  R_{A\to B}(\tau) = D(z_B^\mu(\tau)).
\end{equation}

For detector $B$, one has
\begin{equation}
  \sigma = -{1\over 2}\left[ \rho^2 - UV + {1\over a^2} + {U\over a}e^{a\xi}
  -{2x^0\over a}e^{-a\xi}- {e^{-2a\xi}\over a^2} \right],
\end{equation}
where $U\equiv x^0-x^1$ and $V \equiv x^0+x^1$. So the retarded time of the
field sourced from detector $B$ and observed by detector $A$ at $(t, -d,0,0)$
is $\tau_B^{ret}=\tau(\xi^{ret}(t))$, where
\begin{equation}
  \xi^{ret}(t) = {1\over a} \sinh^{-1}\left[ {a\over 2}(t -d)\right],
\end{equation}
and the retarded distance from detector $B$ to $A$ is
\begin{equation}
  R_{B\to A}(t) = \left|{1\over 2\ell\sqrt{\ell^2+1}}\left[ {2\over a}\ell^3 +
  2t \ell^2 + t+d \right]\right|,
\end{equation}
where $\ell \equiv [a(d-t) + \sqrt{4+ a^2(d-t)^2}]/2$.

\section{Remarks on numerical calculation for correlators}
\label{CalcCorr}

\subsection{Self correlators}

The periodicity of the integrand of~(\ref{dQB2def}) can help to reduce the computation time. For
example, from~(\ref{dQB2def}) one has
\begin{eqnarray}
& &\delta\left<\right.Q_B^2\left(\tau+(\pi/\Omega)\right)\left.\right>_{\rm v} \nonumber\\
&=& {2\gamma\hbar\over \pi m_0\Omega^2}\int_{\tau_0}^{\tau+(\pi/\Omega)}
  d\tilde{\tau}\int_{\tau_0}^{\tau+(\pi/\Omega)}d\tilde{\tau}' K(\tau+(\pi/\Omega)-\tilde{\tau})
  K(\tau+(\pi/\Omega)-\tilde{\tau}')
  \tilde{f}(\tilde{\tau}, \tilde{\tau}')\nonumber\\
&=& e^{-2\pi\gamma/\Omega}\delta\left<\right.Q_B^2(\tau)\left.\right>_{\rm v}
  + {2\gamma\hbar\over \pi m_0\Omega^2}e^{-2\pi\gamma/\Omega}\left[
  \int_{\tau_0}^{\tau}d\tilde{\tau} \int_\tau^{\tau+(\pi/\Omega)}d\tilde{\tau}'+ \right. \nonumber\\ & & 
  \left.\int_\tau^{\tau+(\pi/\Omega)}d\tilde{\tau}\int_{\tau_0}^{\tau}d\tilde{\tau}' +
  \int_\tau^{\tau+(\pi/\Omega)}d\tilde{\tau}\int_\tau^{\tau+(\pi/\Omega)}
  d\tilde{\tau}'\right] K(\tau-\tilde{\tau})K(\tau-\tilde{\tau}')
  \tilde{f}(\tilde{\tau}, \tilde{\tau}'). \nonumber
\end{eqnarray}
Thus one can obtain
$\delta\left<\right.Q_B^2\left(\tau +(\pi/\Omega) \right)\left.\right>_{\rm v}$
by adding the previously obtained
$\delta\left< \right.Q_B^2(\tau)\left.\right>_{\rm v}$ multiplied by a factor
$e^{-2\pi\gamma/ \Omega}$ to the result of an integration over an L-shaped
strip with width $\pi/\Omega$ and total length $2\tau + \pi/\Omega$, rather
than a large $[\tau+(\pi/\Omega)]\times [\tau+(\pi/\Omega)]$ square, in the
$\tilde{\tau} \tilde{\tau}'$-plane. By designing the grid such that there are
exactly $N \in \mathbb{N}$
lattice sites in half a natural period of detector
$\pi/\Omega$ in $\tilde{\tau}$ or $\tilde{\tau}'$, one can improve the
computation time for evaluating
$\delta\left<\right.Q_B^2(\tau)\left.\right>_{\rm v}$ numerically in duration
$\tau_f-\tau_0$ from $O[(\tau_f -\tau_0)^3]$ to
$O[(\tau_f -\tau_0)^2]$~\footnote{We applied Simpson's rule generalized to two
dimensions for numerical integrations in this paper. Each square in our lattice
has five sampling points: the four vertices in the corner and the center point.
Similar to Simpson's $1/3$ rule we assign a $1/12$ weighting factor for the
value of the integrand at each vertex and a $2/3$ factor for the value at the
center. The error would be $O(L^4)$ with lattice constant $L$ and would not
accumulate because the integrand is oscillating.}.

\subsection{Cross correlators }
\label{XcorrRMK}

In~(\ref{XcorrInt}) the non-linear
$t$-dependence of $\tau(t)$ is manifest after $t> 0$, then the domain
$(\tau(t_0), \tau(t))$ that $\tilde{\tau}'$ is integrated over will not
increase in equal time-intervals for each step in $t$. So the trick of
periodicity in obtaining $\delta\left<\right. Q_B^2\left. \right>_{\rm v}^{(0)}$
cannot be applied. However  we can still calculate
$\left<\right. Q_A(t), Q_B(\tau')\left.\right>_{\rm v}^{(0)}$ over the
$t\tau'$-plane, where the periodicity of the integrand can be employed, and
then extract $\left<\right. Q_A(t), Q_B(\tau(t))\left.\right>_{\rm v}^{(0)}$ by
letting $\tau' = \tau(t)$ and interpolating. The results of
$\left<\right. Q_A(t), Q_B(\tau')\left.\right>_{\rm v}^{(0)}$ here is also 
useful in calculating the mutual influences in section~\ref{secmuinf}.

The 
integrand of~(\ref{XcorrInt}) appears to be singular at
$\tilde{\tau}= d+ a^{-1}e^{-a\tilde{\xi}'}$ and
$\tilde{\tau}= d + 2 a^{-1}\cosh a\tilde{\xi}'$ if $\epsilon=0$. However the
presence of the nonzero $\epsilon$ and the fact that the denominators in the
square bracket of the above expression are linear in $\tilde{\tau}$ makes it
possible to deal with the ``singularities'' in the following way:
\begin{eqnarray}
  & &\left<\right. Q_A(t), Q_B(\tau(t))\left.\right>_{\rm v}^{(0)}\nonumber\\
  &=& {2\gamma\hbar\over\pi m_0\Omega^2} {\rm Re} \int_{\tau(t_0)}^{\tau(t)}
    d\tilde{\tau}'\int_{t_0}^t d\tilde{\tau} {K(\tau(t)-\tilde{\tau}')
    \over 2d + a^{-1} e^{a\tilde{\xi}'}}\times\nonumber\\ & &\left[{K(t-\tilde{\tau})-
    K\left(t+d+a^{-1}e^{-a\tilde{\xi}'}\right)+K\left(t+d+a^{-1}e^{-a\tilde{\xi}'}\right)
    \over\tilde{\tau}+d+a^{-1}e^{-a\tilde{\xi}'}-i\epsilon}\right.\nonumber\\ & &\left.
    \hspace{.5cm}-{K(t-\tilde{\tau})-K\left(t-d- 2 a^{-1}\cosh a\tilde{\xi}'\right)
    +K\left(t-d- 2 a^{-1}\cosh a\tilde{\xi}'\right)
    \over \tilde{\tau}-d- 2 a^{-1}\cosh a\tilde{\xi}'-i \epsilon}
    \right]\nonumber
\end{eqnarray}
\begin{eqnarray}                                                                                 
  &=& {2\gamma\hbar\over\pi m_0\Omega^2} {\rm Re} \int_{\tau(t_0)}^{\tau(t)} d\tilde{\tau}'
    {K(\tau(t)-\tilde{\tau}')\over 2d + a^{-1} e^{a\tilde{\xi}'}}
    \times \nonumber\\ & & \left\{\int_{t_0}^t d\tilde{\tau} 
    \left[{K(t-\tilde{\tau})- K\left(t+d+a^{-1}e^{-a\tilde{\xi}'}\right)
    \over\tilde{\tau}+d+a^{-1}e^{-a\tilde{\xi}'}} - \right.\right.\nonumber\\ 
  & & \hspace{2cm}\left. {K(t-\tilde{\tau})-K\left(t-d-2 a^{-1}\cosh a\tilde{\xi}'\right)
    \over \tilde{\tau}-d-2 a^{-1}\cosh a\tilde{\xi}'}\right] \nonumber\\ 
  & & + K\left(t+d+a^{-1}e^{-a\tilde{\xi}'}\right)\ln {\left| t+d+
    a^{-1}e^{-a\tilde{\xi}'}\right|\over \left| t_0+d+a^{-1}e^{-a\tilde{\xi}'}\right|}
    - \nonumber\\ 
  & & \left. K\left(t-d-2 a^{-1}\cosh a\tilde{\xi}'\right) \ln{
    \left|t-d-2 a^{-1}\cosh a\tilde{\xi}'\right|\over
    \left|t_0-d-2 a^{-1}\cosh a\tilde{\xi}'\right|} \right\}.
\label{Xcorrd}
\end{eqnarray}
Those terms in the square bracket of the last expression are smooth, so we can
apply elementary numerical methods such as Simpson's rule to carry out the 2D
integration to high accuracy. The remainder is a one-dimensional integral over
$\tilde{\tau}'$, which is easy to deal with. Although the integrand of the
latter appears to have a logarithmic singularity at
$t_0+d+a^{-1}e^{-a\tilde{\xi}'}=0$, the integral is still finite (and well
defined by $\epsilon$). Note that $|t_0-d-2 a^{-1} \cosh a\tilde{\xi}'|$ is
always positive here, and the combination $K(x)\ln|x|$ is regular at
$|t-d-2 a^{-1}\cosh a\tilde{\xi}'|=0$ and $|t+d+a^{-1}e^{-a\tilde{\xi}'}|=0$.

\begin{figure}
\includegraphics[width=6cm]{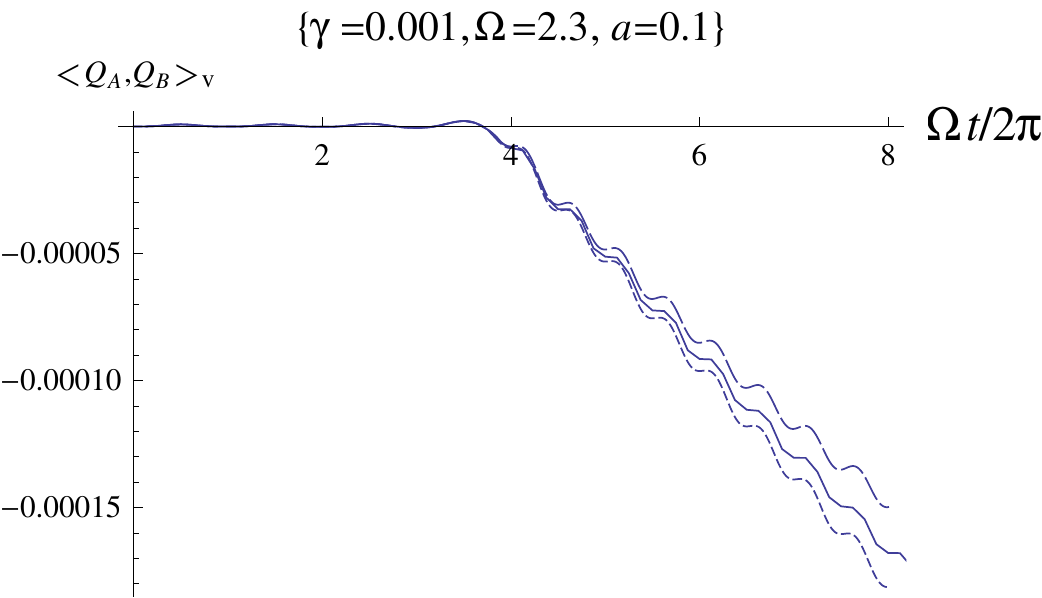}
\includegraphics[width=6cm]{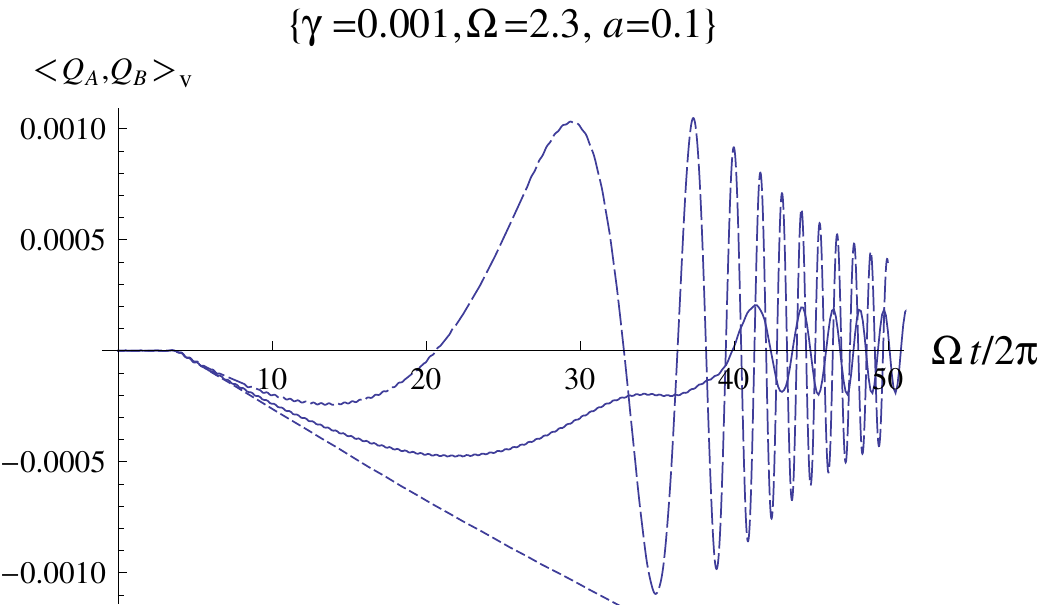}
\caption{Early-time evolution of the cross correlator
$\left<\right. Q_A(t), Q_B(\tau(t)) \left.\right>_{\rm v}$ with $d=10$ 
(solid curves) shown in figure~\ref{XcorrEx} 
compared with the analytical results of the cases with
two inertial detectors sitting at a fixed distance in ref.~\cite{LH09}. The left
plot is a close up of the very early-time behavior in the right plot. The dotted curves
are obtained by inserting the initial distance $d +(t_0+ \sqrt{t_0^2 +(2/a^2)})$
between the two detectors, which is temporally constant, into the
fixed-distance analytical expressions in~\cite{LH09}, while the dashed curves
are obtained by naively inserting the distance $d+(t+ \sqrt{t^2 +(2/a^2)})$ at
each moment $t$ into the same fixed-distance analytic expression. One can see
that at very early times the value of the numerical result here agrees with the
analytic results quite well.} 
\label{Xcomp}
\end{figure}

Comparing the numerical results for the cross correlator
$\left<\right. Q_A(t),Q_B(\tau(t)) \left.\right>_{\rm v}^{(0)}$ and the analytical
results for the case of two inertial detectors sitting at fixed
distance~\cite{LH09} in figure~\ref{Xcomp}, we find excellent agreement at very
early times when the distance between the two detectors is almost constant.

\section{Behavior of 
subtracted self correlator during and after transition}
\label{jump}

The jump of $\delta\left<\right.Q_B^2\left.\right>_{\rm v}$ in figure~\ref{dQB2}
is actually a smooth increase at the same rate as the square of the proper
acceleration~(\ref{avary}) grows. The behavior of
$\delta\left<\right.Q_B^2\left.\right>_{\rm v}$ and the approximately universal
value of the height of the jump  divided by $a^2$ in figure~\ref{dQB2} can be
estimated as follows.

\begin{figure}
\includegraphics[width=7cm]{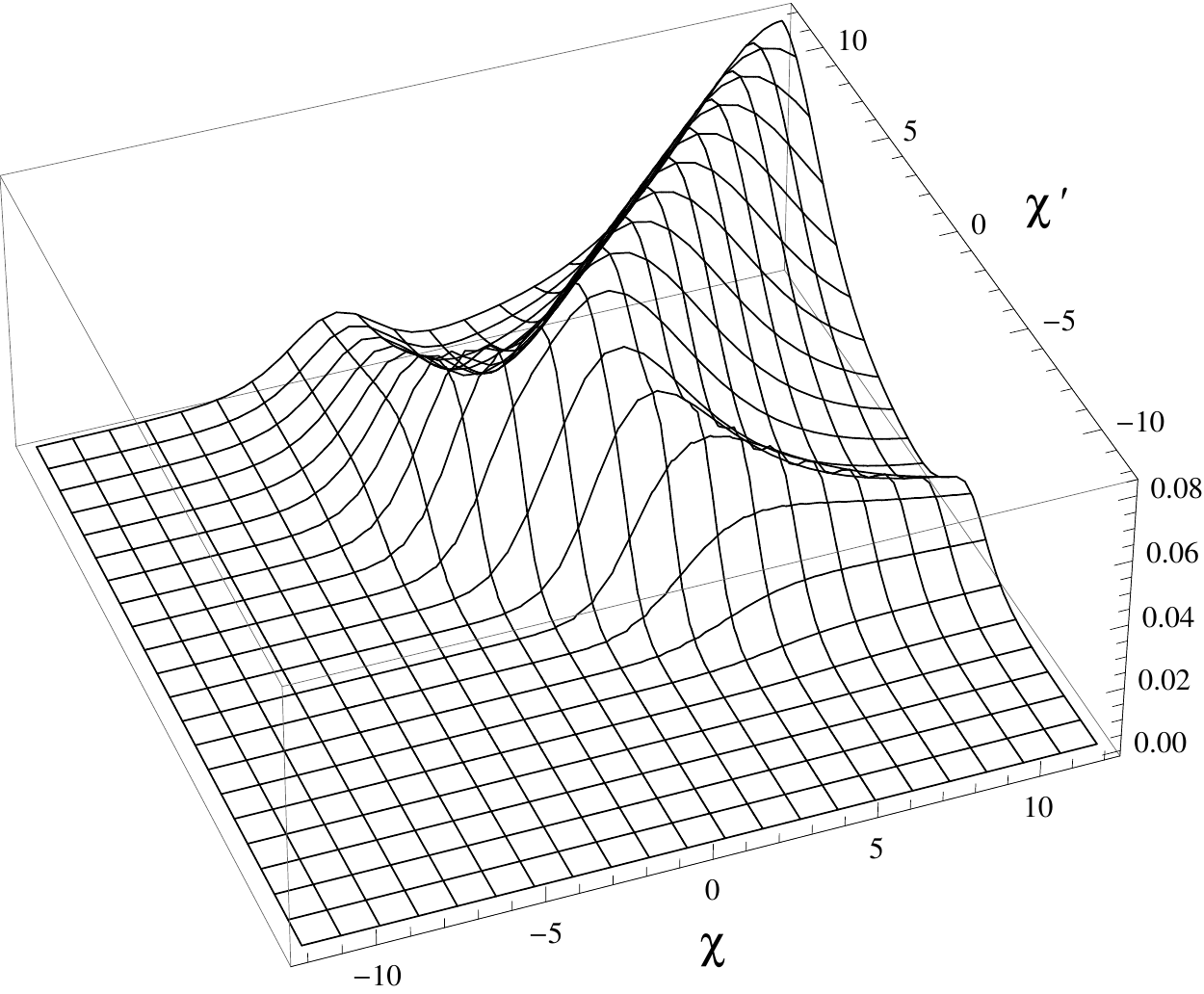}\hspace{.5cm}
\includegraphics[width=6cm]{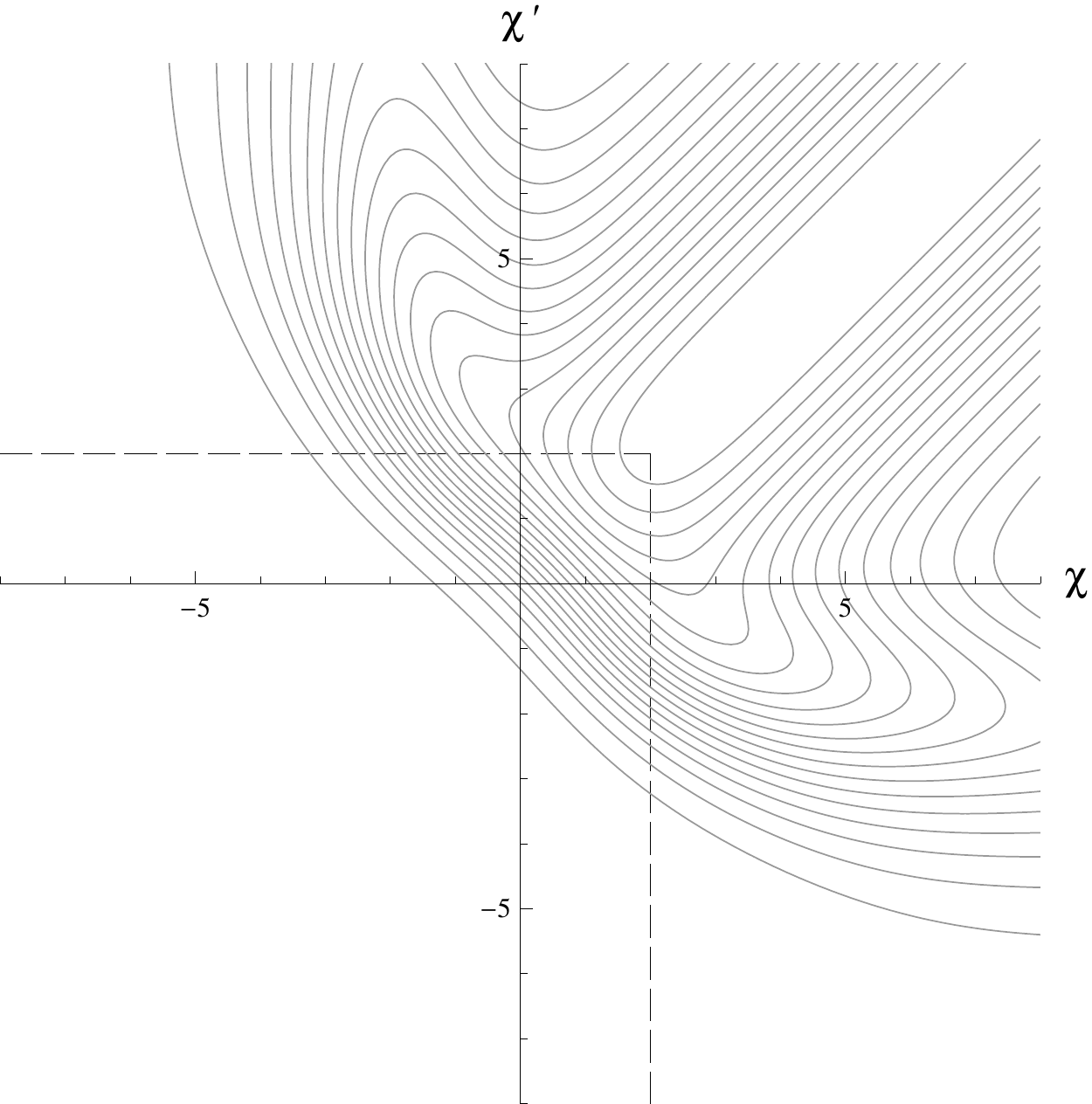}
\caption{The topography (left) and the contour plot (right) of $f(\chi,\chi')$
defined in~(\ref{fdef}).}
\label{fcontour}
\end{figure}

\begin{figure}
\includegraphics[width=6cm]{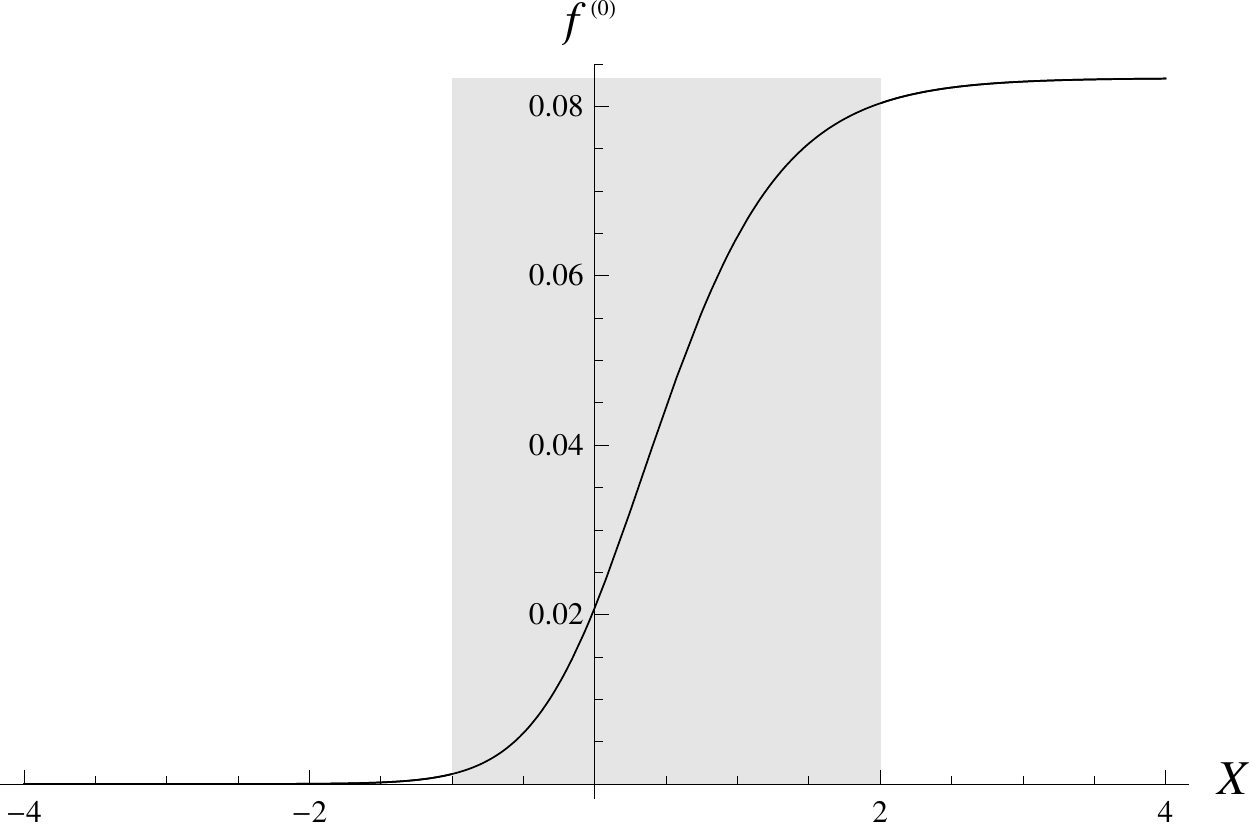}
\includegraphics[width=6cm]{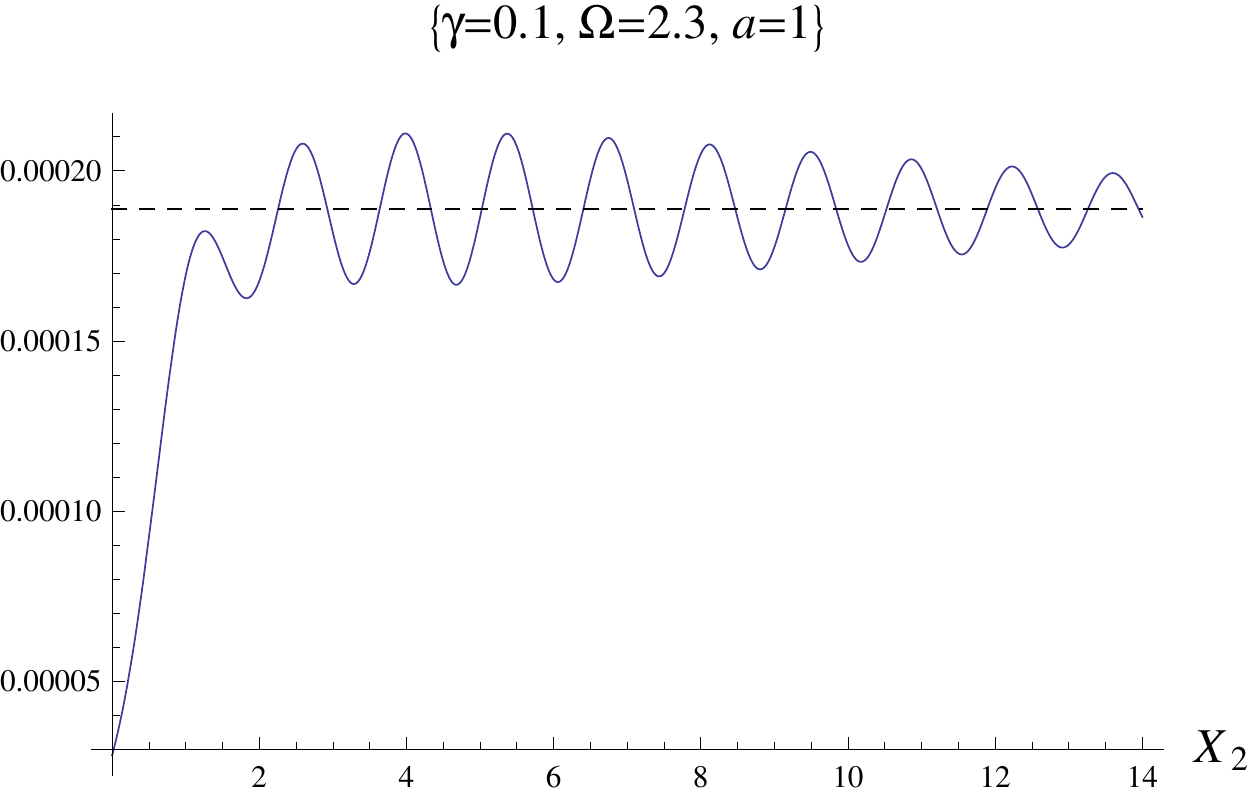}
\caption{(Left) The evolution of the zeroth order term of~(\ref{finD}) in
$X$. The value of $f^{(0)}$ changes significantly around $-1 < X < 2$
(shaded interval). (Right) The values of the 
right hand side of eq.~(\ref{Hyper2F1}) as a function of $X_2$. We see that
 $\left.\delta\left<\right.Q_B^2(\tau)\left.\right>_{\rm v}
\right|_{a\xi(\tau)= X_2}$ grows smoothly from zero after $X_2 \approx 0$, then starts to oscillate
about the limiting value of the function as $X_2\to\infty$.}
\label{f0X}
\end{figure}

The contour plot of $f$ defined in $(\ref{fdef})$ on the $\chi\chi'$-plane is
shown in figure~\ref{fcontour}. One can see that when both $\chi, \chi' < -1$,
the value of $f$ is very small, and when $\chi, \chi' > 0$, a ridge emerges
along $\chi=\chi'$. During the transition $-1 < (\chi, \chi') < 2$,
the values of $f$ in the domain of integration are roughly independent of
$\Delta \equiv \chi-\chi'$ so all the contours are
almost perpendicular to the $X \equiv (\chi+\chi')/2$ directions.
Expanding $f$ in $\Delta$ about $\Delta=0$ yields
\begin{equation}
  f(\chi, \chi') = 
  f^{(0)}(X) 
    - {(1-4e^{-2X})\over 240(1+e^{-2X})^4} \Delta^2 + O(\Delta^4).
\label{finD}
\end{equation}
where 
\begin{equation}
   f^{(0)}(X)\equiv {1\over 12(1+e^{-2X})^2} .
\label{f0Xdef}
\end{equation}
While the zeroth order term of the above expansion $f^{(0)}(X)$
undergoes significant change around $-1 < X < 2$ (see figure~\ref{f0X} (Left)),
the error of the approximation $f(X,\Delta) \approx f^{(0)}(X)$ is always less than $0.05$ 
times of the value of $f(X,\Delta)$ in the region $-1 \le X \le 2$ and $X-2 \le \Delta \le 2-X$.

During the transition $-1\le X \le 2$, the integral in~(\ref{dQB2chi})
is mainly contributed by the integrand in $0 \le X \le 2$, where
$a(\tau-\tau(\xi)) \approx a\xi(\tau) - \chi$ according to~(\ref{TauOfXi}). So
we further approximate
\begin{equation}
  K(\tau-\tau(\chi/a))\approx e^{-\gamma(a\xi(\tau)-\chi)/a}\sin{\Omega\over a}(a\xi(\tau)-\chi).
\label{approx2}
\end{equation}
The error from the deviation of the linearization $a\xi(\tau) - \chi$ from
$a(\tau-\tau(\xi))$ for $X < 0$ will be suppressed efficiently because $f$ is
small while $K(\tau, \tau(\chi/a))$ oscillates wildly there.

Let $X_2(\tau) \equiv a\xi(\tau)$. 
Combining the above approximations for $f\approx f^{(0)}$ in (\ref{f0Xdef}) and $K$ in (\ref{approx2}), and
neglecting the contribution from the $X< -1$ region, we have
\begin{eqnarray}
 & &\delta\left<\right.Q_B^2(\tau)\left.\right>_{\rm v} \nonumber\\
  &\approx& {2\gamma\hbar\over\pi m_0 \Omega^2}
    \int_{\chi(\tau_0)}^{X_2} dX \int_{2(X_2-X)}^{-2(X_2-X)} {d\Delta \over 2} \left[\cos {\Omega\over a}\Delta -
    \cos {2\Omega\over a}(X_2-X) \right]{e^{-2\gamma(X_2-X)/a}\over 12 (1+e^{-2X})^2}\nonumber\\
  &=& {\gamma\hbar\over 12 \pi m_0 \Omega^2} {2a\over \Omega}\left(1-\Omega{\partial\over\partial\Omega}\right)
    {\rm Re}\,\int_{\chi(\tau_0)}^{X_2}dX  {e^{-2(\gamma-i\Omega)(X_2-X)/a}\over i(1+e^{-2X})^2},
\label{dQB2aprx}
\end{eqnarray}
which is expected to be a good approximation for $0 < X_2 < 2$ if $\chi(\tau_0) \ll -1$.
The above integral has an analytic result in a closed form, which is a function of $X_2$. 
For $X_2 \ge 2$ and $\chi(\tau_0)$ is negatively large,
most of the terms goes to zero except the following:
\begin{eqnarray}
  \delta\left<\right.Q_B^2(\tau)\left.\right>_{\rm v}
  &\approx& {\gamma\hbar a \over 12\pi m_0 \Omega^3} \times \nonumber\\ & &
  \left(1-\Omega{\partial\over\partial\Omega}\right){\rm Re}\,\left[ {(1+{\cal W})\over i(2+{\cal W})}
    e^{4X_2} {}_2F_1\left( 1, 2+{\cal W}, 3+{\cal W}, -e^{2X_2}\right)\right],
 \label{Hyper2F1}
\end{eqnarray}
where ${\cal W}\equiv(\gamma+i\Omega)/a$, and ${}_2F_1$ is the hypergeometric function, which oscillates in $X_2$
around a finite, nonzero constant for $X_2 \ge 2$, whose value can be obtained
by, mathematically, taking $X_2 \to \infty$ when the oscillation is damped out
(see figure~\ref{f0X} (Right)).
So we end up with
\begin{eqnarray}
 & & \left. \delta\left<\right.Q_B^2(\tau)\left.\right>_{\rm v}\right|_{a\xi(\tau)=X_2=2}\nonumber\\
 &\approx& {\gamma\hbar a \over 12\pi m_0 \Omega^3}
    \left(1-\Omega{\partial\over\partial\Omega}\right){\rm Re}\,\left[ -ie^{2X_2} +
    {i(1+{\cal W})\over {\cal W}} + O(e^{-2X_2})\right]_{X_2\to\infty}\nonumber\\
    &=& {\gamma\hbar a \over 12\pi m_0 \Omega^3}\left(1-\Omega{\partial\over\partial\Omega}\right){a\Omega\over (\gamma^2+\Omega^2)}
    \nonumber\\ &=&  {\gamma\hbar a^2 \over 6\pi m_0 (\gamma^2+\Omega^2)^2},
\end{eqnarray}
which is consistent with our observations in figure~\ref{dQB2} and eq.~(\ref{dQBjump}).

Now the numerical behavior of
$\delta\left<\right.Q_B^2(\tau)\left.\right>_{\rm v}$ can be understood as
follows. The domain of the $\chi\chi'$-integration is the square with both
$\chi$, $\chi'\in [a\xi(\tau_0), a\xi(\tau)]$. $X_2(\tau) \equiv a\xi(\tau)$
increases as $\tau$ increases. When the vertex $(X_2(\tau), X_2(\tau))$ 
of the domain touches $X\approx -1$, the growth of $f(\chi,\chi)$ becomes significant so
$\delta\left<\right.Q_B^2(\tau)\left.\right>_{\rm v}$ starts to grow. The
latter keeps growing smoothly until the vertex of the domain reaches
$X\approx 2$ (the boundary of the domain at this moment is indicated by the
dashed lines in figure~\ref{fcontour} (Right)), then the evolution enters another
phase where $\delta\left<\right.Q_B^2(\tau)\left.\right>_{\rm v}$ grows slowly
in a time scale of $1/2\gamma$ as shown in~(\ref{evoweak}), with
oscillations on top of the growing, towards the late-time value
$\delta\left<\right.Q_B^2(\infty)\left.\right>_{\rm v}$.

\begin{figure}
\includegraphics[width=6cm]{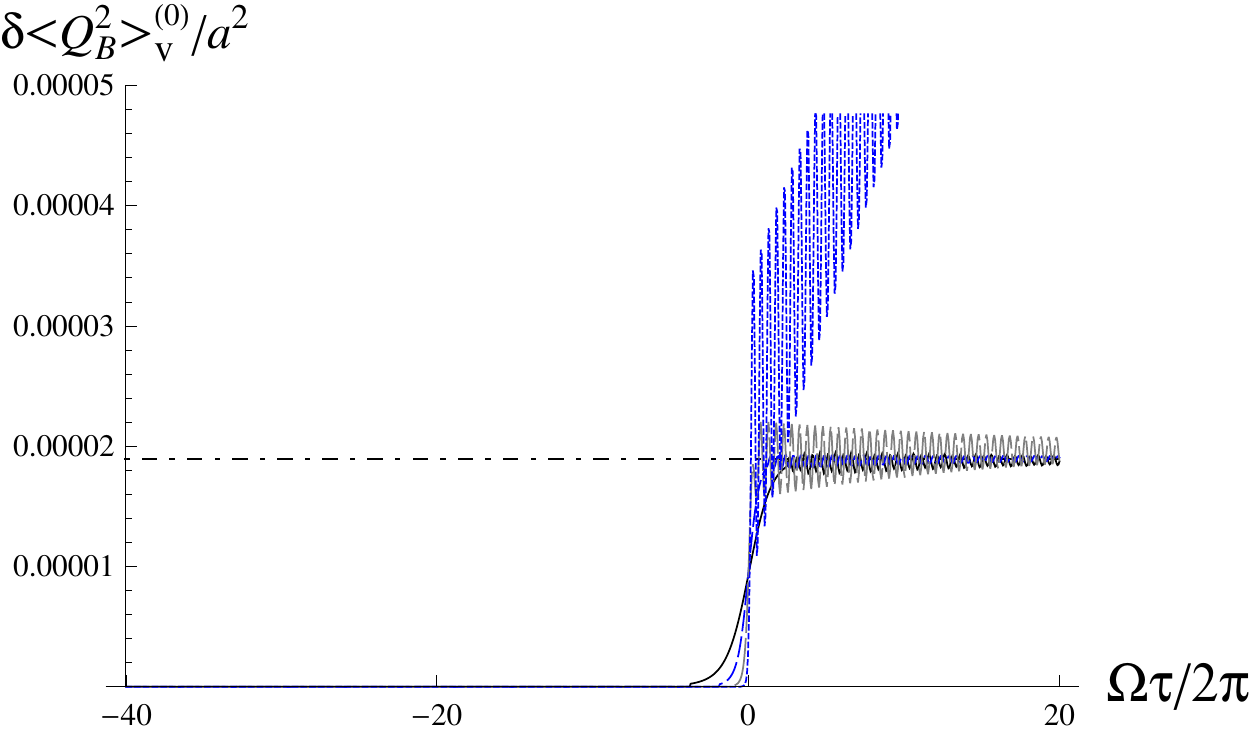}
\includegraphics[width=6cm]{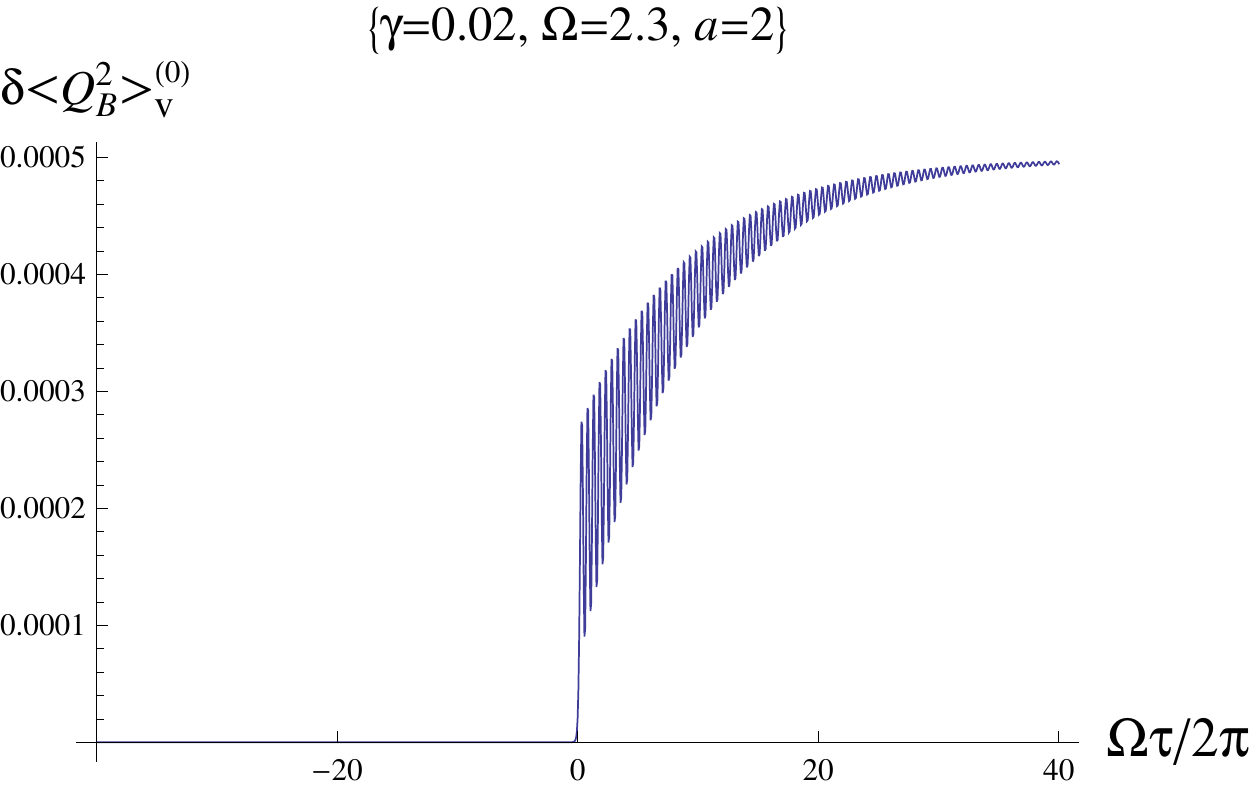}
\caption{Approximated evolution of $\delta\left<\right. Q_B^2(\tau)\left.\right>_{\rm v}$
by $(\ref{ourapprox})$. The left and the right plots are made for comparison with
the lower-left plot of Figure \ref{dQB2} and the left plot of Figure \ref{alarge}.
Our approximation agrees well with the numerical results qualitatively, except
the non-adiabatic oscillations are over-estimated. The horizontal dot-dashed line
in the left plot indicates the value of ${\cal Q}/a^2$ with ${\cal Q}$ defined in $(\ref{dQBjump})$.}
\label{QBvapprox}
\end{figure}

\begin{figure}
\includegraphics[width=6cm]{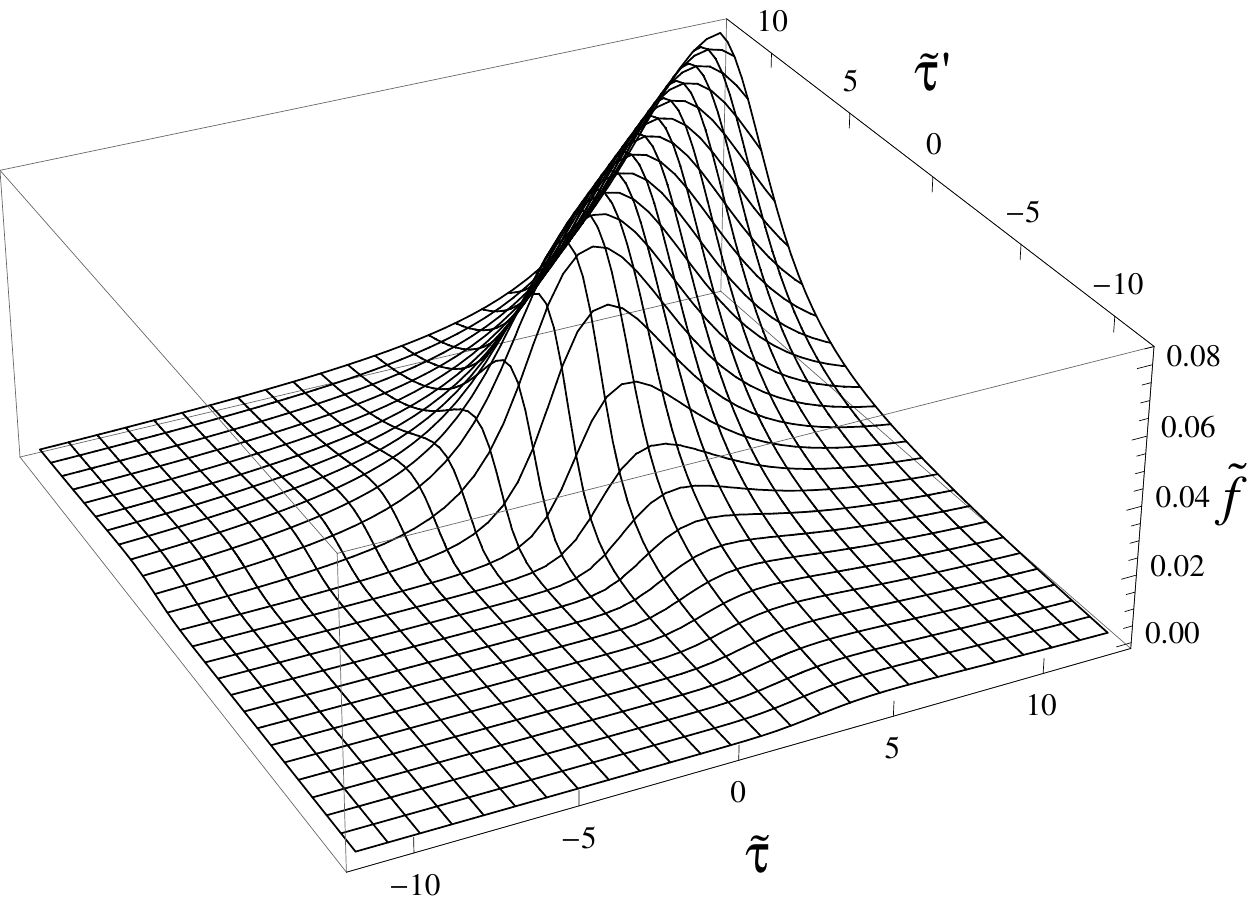}
\includegraphics[width=6.7cm]{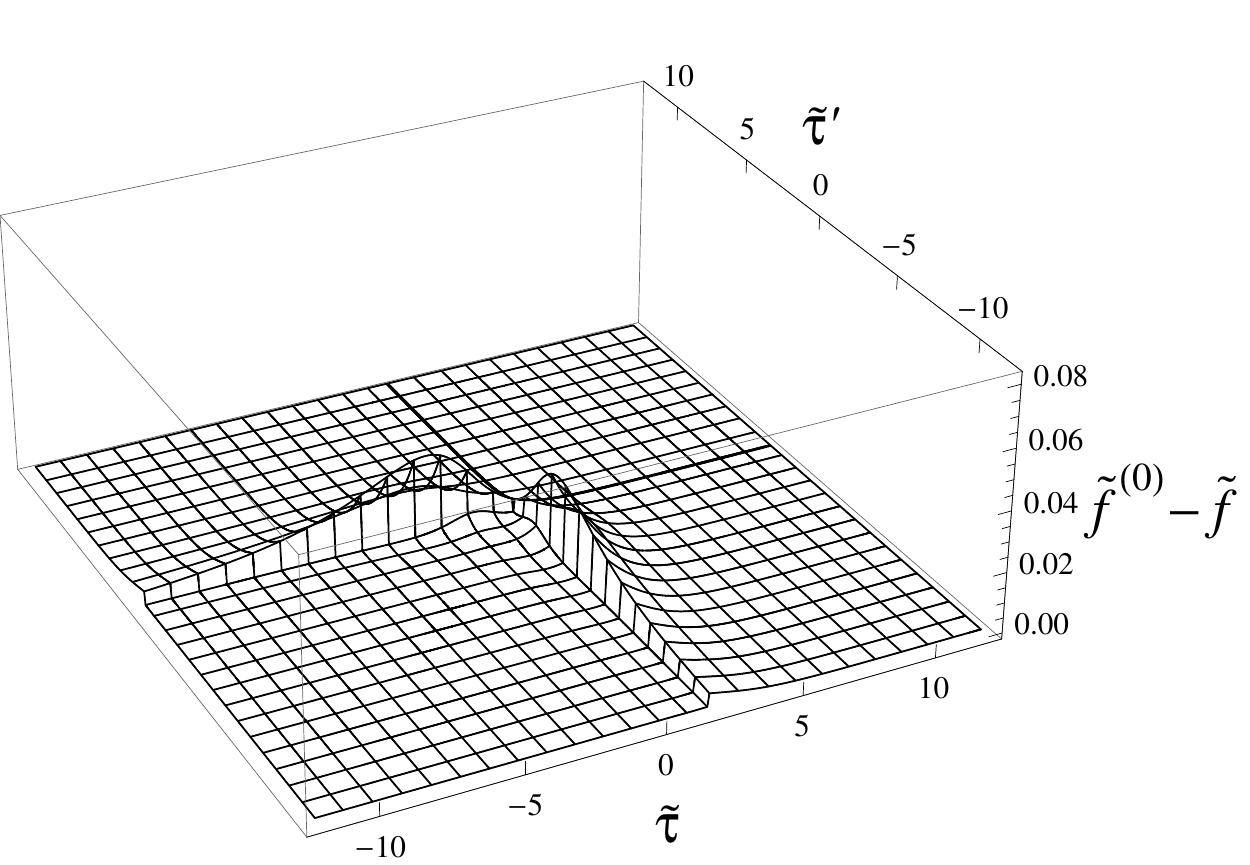}
\caption{A comparison between $\tilde{f}$ given in $(\ref{tildefdef})$ (left)
and the error of our approximation $\tilde{f}^{(0)}-\tilde{f}$ (right) on the
$\tilde{\tau}\tilde{\tau}'$-plane. Here $a=1$, $\tilde{f}^{(0)} = f^{(0)}(X(\tilde{\tau},\tilde{\tau}'))$
defined in $(\ref{f0Xdef})$ for $\tilde{\tau}, \tilde{\tau}' < \tau_2$, and
$\tilde{f}^{(0)}(\tilde{\tau},\tilde{\tau}')$ defined in $(\ref{tildef})$ otherwise.
One can see that the error is quite localized and mostly positive.}
\label{fapprox}
\end{figure}

To obtain more insight into the behavior of $\delta\left<\right.Q_B^2(\tau)\left.\right>_{\rm v}$
after $X_2 \approx 2$, let us consider the following simple approximations.
Let $ \tau^{}_{2}\equiv \tau(\xi)|_{a\xi=2}$. In the region $\tau^{}_{2} \le \tilde{\tau}
\le \tau$ or $ \tau^{}_{2} \le \tilde{\tau}'\le \tau$ we observed that
\begin{equation}
  \tilde{f}(\tilde{\tau},\tilde{\tau}') \approx  \tilde{f}^{(0)}(\tilde{\Delta}) \equiv
  {1\over \tilde{\Delta}^2} - {a^2\over 4\sinh^2 (a\tilde{\Delta}/2)} 
\label{tildef}
\end{equation}
in the integrand of $(\ref{dQB2def})$ with $\tilde{\Delta}\equiv \tilde{\tau}'-\tilde{\tau}$ .
So $\delta\left<\right.Q_B^2(\tau)\left.\right>_{\rm v}$ for $\tau> \tau^{}_{2}\gg \tau_0$ can be
approximated by
\begin{equation}
  \delta\left<\right.Q_B^2(\tau)\left.\right>_{\rm v} \approx
  \delta\left<\right.Q_B^2(\tau)\left.\right>_{\rm v}^{(0)} \equiv
  {2\gamma\hbar\over\pi m_0 \Omega^2} \left[ I_1(\tau) + I_2(\tau) \right],
\label{ourapprox}
\end{equation}
where
\begin{eqnarray}
  I_1 &\equiv&
    \int_{\chi(\tau_0)}^{2}d\chi \int_{\chi(\tau_0)}^{2}d\chi'
    K\left(\tau-\tau(\chi/a)\right)K\left(\tau-\tau(\chi'/a)\right)f^{(0)}(X), \nonumber\\
  I_2 &\equiv& \left( \int_{\tau_0}^{\tau}d\tilde{\tau}\int_{\tau_0}^{\tau}d\tilde{\tau}' -
    \int_{\tau_0}^{ \tau^{}_{2}}d\tilde{\tau}\int_{\tau_0}^{ \tau^{}_{2}}d\tilde{\tau}'\right)
    K(\tau-\tilde{\tau})K(\tau-\tilde{\tau}')\tilde{f}^{(0)}(\tilde{\Delta}).
\end{eqnarray}
By modifying the earlier calculation in obtaining $(\ref{Hyper2F1})$, it is straightforward to see
\begin{equation}
  I_1 \approx {e^{-2\gamma\eta +4\times 2}\over 24\Omega} 
  {\rm Re}\left\{\left( 1-e^{-2i\Omega\eta}\Omega{\partial\over\partial\Omega}\right)
  \left[ {a\over i}\left({1+{\cal W}\over 2+{\cal W}}\right)
  F_{1+{\cal W}}(-e^{2\times 2}) \right] \right\}
\end{equation}
with $F_y(x) \equiv {}_2F_1(1+y,1,2+y,x)$, and $\eta \equiv \tau - \tau^{}_{2}$ since $\tau \sim \xi \, (= \chi/a)$
in this region by $(\ref{TauOfXi})$. For $I_2$, by noting that the first and the second terms of $\tilde{f}^{(0)}$
in $(\ref{tildef})$ are nothing but the Hadamard functions of the massless scalar field experienced
by an inertial detector and a uniformly accelerated detector with proper acceleration $a$, respectively
\cite{BD82, LH06}, we apply the techniques similar to those in Refs. \cite{LH06} and \cite{LH07} to obtain
\begin{eqnarray}
   I_2 &=& {\pi m_0 \Omega^2\over2\gamma\hbar}\left[
   \left.\left<\right.Q^2(\eta^{}_0)\left.\right>_{\rm v}\right|_{a_\mu a^\mu = a^2} -
   \left.\left<\right.Q^2(\eta^{}_0)\left.\right>_{\rm v}\right|_{a_\mu a^\mu \to 0}\right] 
 \nonumber\\ && -{1\over 2}{\rm Re}\left\{ e^{-2\gamma\eta}
  \left( 1- e^{-2i\Omega\eta} + {i\Omega\over\gamma}\right) \times \right.\nonumber\\
   & & \hspace{1cm} \left. \left[
   \Gamma\left(0, {\cal W} a\bar{\eta} \right) -
   {e^{-(1+{\cal W})a\bar{\eta}} \over 1+{\cal W}} F_{{\cal W}}\left( e^{-a \bar{\eta}} \right)
   - \psi\left(1+{\cal W}\right) + {1\over 2{\cal W}} + \ln {\cal W} \right] +\right. \nonumber\\
   && e^{-2\gamma\eta^{}_0}
  \left( 1- e^{-2i\Omega\eta^{}_0} + {i\Omega\over\gamma}\right)\times \nonumber\\ & & \left.\left[
   \Gamma\left(0, -{\cal W} a\bar{\eta} \right) -
   {e^{-(1-{\cal W})a\bar{\eta}} \over 1-{\cal W}} F_{-{\cal W}}\left( e^{-a \bar{\eta}} \right)
   - \psi\left(1-{\cal W}\right) - {1\over 2{\cal W}}  + \ln (-{\cal W}) \right]\right\}
\end{eqnarray}
with $\eta^{}_0 \equiv \tau-\tau_0$, $\bar{\eta}\equiv \tau^{}_{2}-\tau_0$, 
and the v-part of the self correlator $\left.\left<\right.Q^2(\tau-\tau_0)\left.\right>_{\rm v}
\right|_{a_\mu a^\mu = a^2}$ of a uniformly accelerated detector with proper acceleration $a$ moving
in a massless scalar field initially in vacuum state (see eqs. (A3) and (A9) in \cite{LH07}).
In figure \ref{QBvapprox} we illustrate that the above approximation can indeed describe the behavior after
$\tau > \tau^{}_{2}$ qualitatively.
The major difference is the amplitude of the non-adiabatic oscillations on top of the rising curve.
Since the $1/\tilde{\Delta}^2$ term in $\tilde{f}$ or $\tilde{f}^{(0)}$ in $(\ref{tildef})$ dominates
whenever $|\tilde{\Delta}|$ is large,
the error of the above approximation will be localized in the vicinity of ($\tilde{\tau}\approx  \tau^{}_{2}$,
$\tilde{\tau}' <  \tau^{}_{2}$) and ($\tilde{\tau} <  \tau^{}_{2}$, $\tilde{\tau}'\approx  \tau^{}_{2}$)
with small $\tilde{\tau}-\tilde{\tau}'$.
As shown in figure \ref{fapprox} (right), $\tilde{f}^{(0)}-\tilde{f}$ is mostly positive,
so the approximation $(\ref{ourapprox})$ usually gives the
non-adiabatic oscillations a larger amplitude than the true amplitude, 
while these oscillations will be
damped out at late times. In the weak coupling limit, the approximation $(\ref{ourapprox})$
behaves similarly to $(\ref{evoweak})$.

The behavior of $\delta\left<\right.Q_B, P_B\left.\right>_{\rm v}$ during the
transition can be obtained straightforwardly since
$\left<\right.Q_B(\tau), P_B(\tau)\left.\right>_{\rm v}
= \partial_\tau \left<\right.Q_B^2(\tau) \left.\right>_{\rm v} /2$. For
$\delta\left<\right. P_B^2\left.\right>$, the calculation is similar except
that the functions $K(\tau -\tau(\chi/a))$ in~(\ref{dQB2chi}) are replaced by
\begin{eqnarray}
  K'(\tau -\tau(\chi/a)) &=& e^{-\gamma(\tau -\tau(\chi/a))}\left[\Omega \cos\Omega\left(\tau -\tau(\chi/a)\right)
    -\gamma \sin\Omega\right(\tau -\tau(\chi/a)\left) \right] \nonumber\\
  &\approx& e^{-\gamma(a\xi(\tau)-\chi)/a}\left[ \Omega \cos{\Omega\over a}\left(a\xi(\tau)-\chi \right)
    -\gamma \sin{\Omega\over a}\left( a\xi(\tau)-\chi\right) \right]
\end{eqnarray}
during $-1 < X < 2$. This gives
$\left.\delta\left<\right.P_B^2(\tau)\left.\right>_{\rm v}\right|_{a\xi(\tau)=X_2} \approx 0$ as $X_2 \to \infty$,
which is consistent with the observations in Figs.~\ref{weakag} and~\ref{alarge}
that there is no significant jump for
$\delta\left<\right.P_B^2(\tau)\left.\right>_{\rm v}$ around $\tau\approx 0$.

Suppose $a(\tau_2-\tau_0) \gg 1$. Then the amplitude of the non-abiabatic oscillations in (\ref{ourapprox})
will be
\begin{eqnarray}
&& {\gamma\hbar e^{-2\gamma\eta} \over\pi m_0 \Omega^2} \left| {iae^{8}\over 12} {\partial\over\partial\Omega}
  \left[\left({1+{\cal W}\over 2+{\cal W}}\right)F_{1+{\cal W}}(-e^{4})
  \right] + {1\over 2{\cal W}} + \ln {\cal W} - \psi\left(1+{\cal W}\right)  \right| \nonumber\\  
&\approx& {\gamma\hbar e^{-2\gamma\eta} \over 2\pi m_0 \Omega^2} {a\over \sqrt{\Omega^2+\gamma^2}}
\end{eqnarray}
for $a \gg \sqrt{\Omega^2+\gamma^2}$. This is approximately the amplitude for the cases with a sudden rise of proper 
acceleration from $0$ to $a$. Thus if detector $B$ has been almost in the steady state before $\tau_2$, 
the amplitude of the non-adiabatic oscillations of the values of its self correlators after $\tau_2$ will be no greater 
than $O(\gamma a)$ for large $a$. 
Since the values of the self correlators of detector $B$ are $O(\gamma^0 a)$ at late times \cite{LH07},
those non-adiabatic oscillations will not be significant in the ultraweak coupling limit.
Of course if $a(\tau_2-\tau_0)$ is not very large and so detector $B$ is far from steady state at $\tau_2$, 
the non-adiabatic oscillations can be enhanced.

\end{appendix}


\begin{thebibliography}{9}

\bibitem{Un76} W. G. Unruh, {\it Notes on black-hole evaporation, Phys. Rev. D} {\bf 14} (1976) 870.


\bibitem{AM03} P. M. Alsing and G. J. Milburn, {\it Teleportation with a uniformly accelerated partner,
Phys. Rev. Lett.} {\bf 91} (2003) 180404 [quant-ph/0302179].

\bibitem{SU05} R. Sch\"utzhold and W. G. Unruh, {\it Comment on 
``Teleportation with a uniformly accelerated partner"}, preprint [quant-ph/0506028].

\bibitem{FueMan05} I. Fuentes-Schuller and R. B. Mann, {\it Alice falls into 
a black hole: Entanglement in non-inertial frames, Phys. Rev. Lett.}
{\bf 95} (2005) 120404 [quant-ph/0410172].

\bibitem{FueMan06} P. M. Alsing, I. Fuentes-Schuller, R. B. Mann, and T. E.
Tessier, {\it Entanglement of Dirac fields in non-inertial frames, Phys. Rev. A} 
{\bf 74} (2006) 032326 [quant-ph/0603269].

\bibitem{OPhD08} D. C. M. Ostapchuk, {\it Entanglement in Non-inertial
Frames}, Master thesis, University of Waterloo (2008).

\bibitem{OM09} D. C. M. Ostapchuk and R. B. Mann, {\it Generating entangled fermions 
by accelerated measurements on the vacuum, Phys. Rev. A} {\bf 79} (2009) 042333 [arXiv:0903.0219].

\bibitem{LH09} S.-Y. Lin and B. L. Hu, {\it Temporal and spatial dependence of quantum entanglement 
from a field theory perspective, Phys. Rev. D} {\bf 79} (2009) 085020 [arXiv:0812.4391].

\bibitem{LCH08} S.-Y. Lin, C.-H. Chou, and B. L. Hu, {\it Disentanglement of two
harmonic oscillators in relativistic motion, Phys. Rev. D} {\bf 78}
(2008) 125025 [arXiv:0803.3995].

\bibitem{LH10} S.-Y. Lin and B. L. Hu, {\it Entanglement creation between two causally 
disconnected objects, Phys. Rev. D} {\bf 81} (2010) 045019 [arXiv:0910.5858].

\bibitem{RHK} A. Raval, B. L. Hu and D. Koks, {\it Near-thermal radiation in detectors, 
mirrors and black holes: a stochastic approach, Phys. Rev. D} {\bf 55} (1997) 4795 [gr-qc/9606074].

\bibitem{JHCapri} B. L. Hu, Philip R. Johnson, {\it Beyond Unruh Effect:
Nonequilibrium Quantum Dynamics of Moving Charges}, Invited Talk at the Capri
Workshop on Quantum spects of Beam Physics, Oct. 2000. Proceedings edited by
Pisin Chen (World-Scientific, Singapore, 2001) [quant-ph/0012132].

\bibitem{BL83} J. S. Bell and J. M. Leinaas, {\it Electrons as accelerated thermometers,
Nucl. Phys. B} {\bf 212}, 131 (1983) 131;
J. S. Bell and J. M. Leinaas, {\it The Unruh effect and quantum fluctuations of electrons in storage rings,
Nucl. Phys. B} {\bf 284} (1987) 488;
W. G. Unruh, {\it Acceleration radiation for orbiting electrons, Phys. Rep.} {\bf 307} (1998) 163 [hep-th/9804158];
E. T. Akhmedov, and D. Singleton, {\it On the relation between Unruh and Sokolov--Ternov effects,
Int. J. Mod. Phys.} {\bf A22} (2007) 4797 [hep-th/0610391];
E. T. Akhmedov and D. Singleton, {\it On the physical meaning of the Unruh effect,
Pisma Zh. Eksp. Teor. Fiz.} {\bf 86} (2007) 702 [arXiv:0705.2525].

\bibitem{SS92} B. F. Svaiter and N. F. Svaiter, {\it Inertial and noninertial particle 
detectors and vacuum fluctuations, Phys. Rev. D}
{\bf 46} (1992) 5267; {\it Erratum ibid} {\bf 47} (1993) 4802.

\bibitem{HMP93} A. Higuchi, G. E. A. Matsas and C. B. Peres,
{\it Uniformly accelerated finite-time detectors, Phys. Rev. D} {\bf 48} (1993) 3731.

\bibitem{Pad96} L. Sriramkumar and T.Padmanabhan, {\it Response of finite-time particle detectors in non-inertial 
frames and curved spacetime, Class. Quantum Grav.} {\bf 13} (1996) 2061 [gr-qc/9408037].

\bibitem{OP03} N. Obadia and R. Parentani, {\it Uniformly accelerated mirrors. 
Part 1: Mean fluxes, Phys. Rev. D} {\bf 67} (2003) 024021 [gr-qc/0208019]; 
{\it Uniformly accelerated mirrors. Part 2: Quantum correlations, ibid} 
{\bf 67} (2003) 024022 [gr-qc/0209057].

\bibitem{Sc04} S. Schlicht, {\it Considerations on the Unruh Effect: Causality and Regularization,
Class. Quantum Grav.} {\bf 21} (2004) 4647 [gr-qc/0306022];
J. Louko and A. Satz, {\it How often does the Unruh-DeWitt detector click? Regularisation by a spatial profile,
Class. Quantum Grav.} {\bf 23} (2006) 6321 [gr-qc/0606067];
P. Langlois, {\it Causal particle detectors and topology, Annals Phys.} {\bf 321} (2006) 2027 [gr-qc/0510049].

\bibitem{OM07} N. Obadia and M. Milgrom, {\it Unruh effect for general trajectories,
 Phys. Rev. D} {\bf 75} (2007) 065006 [gr-qc/0701130]. 

\bibitem{CV99} R. Casadio and G. Venturi, {\it The accelerated observer with back-reaction effects,
Phys. Lett.} {\bf A252} (1999) 109 [gr-qc/9810073].

\bibitem{KP10} D. Kothawala and T. Padmanabhan, {\it Response of Unruh-DeWitt detector with time-dependent acceleration,
Phys. Lett.} {\bf B 690} (2010) 201 [arXiv:0911.1017].

\bibitem{LH07} S.-Y. Lin and B. L. Hu, {\it Backreaction and the Unruh effect:
New insights from exact solutions of uniformly accelerated detectors, Phys. Rev. D} {\bf 76} 
(2007) 064008 [gr-qc/0611062].

\bibitem{CalHu08} E. A. Calzetta  and B.-L.  Hu, {\it Nonequilibrium Quantum
Field Theory}, Cambridge University Press, Cambridge U.K. (2008).

\bibitem{Kalnins} E. G. Kalnins, {\it On the separation of variables for the Laplace equation
$\Delta\psi + K^2 \psi =0$ in two- and three-dimensional Minkowski space, SIAM J. Math. Anal.} 
{\bf 6} (1975) 340.

\bibitem{CV} I. Costa, {\it Separable coordinates and particle creation. I: 
the Klein-Gordon equation, Rev. Bras. Fis.} {\bf 17} (1987) 585;
I. Costa, {\it Separable coordinates and particle creation. II: Two new vacua related to 
accelerating observers, J. Math. Phys.} {\bf 30} (1989) 888;
I. Costa and N. F. Svaiter, {\it Separable coordinates and particle creation. III: 
Accelerating, Rindler, and Milne vacua, Rev. Bras. Fis.} {\bf 19} (1989) 271.

\bibitem{PV92} U. Percoco and V. M. Villalba, {\it Particle creation in an asymptotically uniformly 
accelerated frame, Class. Quantum Grav.} {\bf 9} (1992) 307.

\bibitem{VM92}V. M. Villalba and J. Mateu, {\it Vacuum effects in an asymptotically uniformly 
accelerated frame with a constant magnetic field, Phys. Rev. D} {\bf 61} (2000) 025007
[hep-th/9910072].

\bibitem{MV09} R. B. Mann and V. M. Villalba, {\it Speeding up Entanglement Degradation,
Phys. Rev. A} {\bf 80} (2009) 022305 [arXiv:0902.1580].

\bibitem{LinBehHu} S.-Y. Lin, R. Behunin and B. L. Hu, {\it Quantum Twin Paradox:
Entanglement is memory-laden}, in preparation.

\bibitem{LH06} S.-Y. Lin and B. L. Hu, {\it Accelerated detector - quantum field correlations:
From vacuum fluctuations to radiation flux, Phys. Rev. D} {\bf 73} (2006) 124018 [gr-qc/0507054].

\bibitem{BD82} N. D. Birrell and P. C. W. Davies, {\it Quantum Fields in
Curved Space}, Cambridge University Press, Cambridge U.K. (1982).

\bibitem{Si00} R. Simon, {\it Peres-Horodecki separability criterion for continuous variable systems,
Phys. Rev. Lett.} {\bf 84} (2000) 2726 [quant-ph/9909044]; 
L.-M. Duan, G. Giedke, J. I. Cirac, and P. Zoller, {\it Inseparability Criterion for 
Continuous Variable Systems, Phys. Rev. Lett.} {\bf 84} (2000) 2722 [quant-ph/9908056].

\bibitem{VW02} G. Vidal and R. F. Werner, {\it Computable measure of entanglement,
A computable measure of entanglement, Phys. Rev. A} {\bf 65} (2002) 032314 [quant-ph/0102117].

\bibitem{cosmo}J. L. Ball, I. Fuentes-Schuller, and F. P. Schuller, {\it Entanglement in an expanding spacetime,
Phys. Lett.} {\bf A359} (2006) 550 [quant-ph/0506113]; 
G. L. Ver Steeg and N. C. Menicucci, {\it Entangling Power of an Expanding Universe,
Phys. Rev. D} {\bf 79} (2009) 044027 [arXiv:0711.3066]; 
I. Fuentes, R.B. Mann,  E. Martin-Martinez, and S. Moradi, {\it Entanglement of 
Dirac fields in an expanding spacetime, Phys. Rev. D} {\bf 82} (2010) 045030 [arXiv:1007.1569].
\bibitem{Wilczek} M. K. Parikh and F. Wilczek, {\it An action for black hole membranes, Phys. Rev. D} {\bf 58}
(1998) 064011 [gr-qc/9712077].

\bibitem{Hayward} S. A. Hayward, {\it General laws of black-hole dynamics, Phys. Rev. D} {\bf 49} (1994) 6467
[gr-qc/9303006].

\bibitem{Ash} A. Ashtekar and B. Krishnan, {\it Isolated and Dynamical Horizons and
Their Applications}, Living Rev. Relativity {\bf 7} (2004) 10.
URL (cited on 25 July 2010): http://www.livingreviews.org/lrr-2004-10

\bibitem{LSCH12} S.-Y. Lin, K. Shiokawa, C.-H. Chou, and B. L. Hu, 
{\it Quantum teleportation between moving detectors in a quantum field} [arXiv:1204.1525].

\end{thebibliography}
\end{document}